# Efficient photoreforming of plastic waste using a high-entropy oxide catalyst

Thanh Tam Nguyen[a,b] and Kaveh Edalati[a,b,*]

[a] WPI, International Institute for Carbon Neutral Energy Research (WPI-I2CNER), Kyushu University, Fukuoka 819-0395, Japan

[b] Mitsui Chemicals, Inc. - Carbon Neutral Research Center (MCI-CNRC), Kyushu University, Fukuoka 819-0395, Japan

Simultaneous catalytic hydrogen ($H_2$) production and plastic waste degradation under light, known as photoreforming, is a novel approach to green fuel production and efficient waste management. Here, we use a high-entropy oxide (HEO), a new family of catalysts with five or more principal cations in their structure, for plastic degradation and simultaneous $H_2$ production. The HEO shows higher activity than that of P25 $TiO_2$, a benchmark photocatalyst, for the degradation of polyethylene terephthalate (PET) plastics in water. Several valuable products are produced by photoreforming of PET bottles and microplastics including $H_2$, terephthalate, ethylene glycol and formic acid. The high activity is attributed to the diverse existence of several cations in the HEO lattice, lattice defects, and appropriate charge carrier lifetime. These findings suggest that HEOs possess high potential as new catalysts for concurrent plastic waste conversion and clean $H_2$ production.

*Corresponding author (E-mail: kaveh.edalati@kyudai.jp; Tel: +80-92-802-6744)

1.  **Introduction**

Plastics have become an indispensable part of modern society for a vast range of applications including the food industry, agriculture, and medical systems [1]. The widespread use of plastics is due to numerous advantages such as low density, versatility, and resistance to corrosion and fracture. However, a significant environmental challenge arises from plastic waste, with over 80% of plastic packages ending up in landfills or polluting the environment [2]. In addition to environmental problems, they can also cause a huge loss of valuable resources. Small pieces of plastic that are less than 5 mm in size are usually referred to as microplastics. Microplastics have become pervasive and abundant in various environments such as drinking water, oceans, terrestrial lands, and the atmosphere. Nowadays, microplastics are recognized as emergent environmental hazards and have drawn a lot of community attention due to their severe toxicity to ecosystems and human beings. It is extremely challenging to gather and reuse microplastics because of their small size and dilution in the environment. The most prevalent polyester plastic and a significant contributor to plastic waste is polyethylene terephthalate (PET), which is basically recyclable. However, although around 30 million tons of PET are globally produced, only 9% of PET is practically recycled [3]. Current waste management systems are unable to efficiently and affordably treat a variety of plastic waste including PET microplastics, and thus, new technologies are urgently needed to convert microplastic waste into useful products. Photocatalysis is a technology that has been recently developed for microplastic treatment to produce valuable organic chemicals and fuels [4,5].

In addition to environmental pollution by microplastics, the world is also facing an energy crisis and $CO_2$ emissions by fossil fuels. It is important to develop sustainable fuels and novel technologies to effectively tackle this energy crisis. Hydrogen ($H_2$), characterized by its eco-friendly nature, presents itself as an ideal clean energy carrier [6,7]. $H_2$ is primarily produced through energy-intensive and unclean processes. For instance, the steam-methane reforming method, which is used worldwide, necessitates the use of high-temperature (973-1273 K) and high-pressure (3-25 bar) [7]. Moreover, a significant amount of $CO_2$ is emitted into the atmosphere during steam reforming. Photocatalytic $H_2$ production from water has emerged as an effective and clean alternative to produce $H_2$ at ambient temperature and pressure utilizing solar energy as a power source. In this process, photoexcited electrons of the catalyst contribute to $H_2$ production and corresponding holes contribute to $O_2$ production. However, limited catalysts can produce both $H_2$ and $O_2$, and thus, most of the $H_2$ production systems need a sacrificial agent for holes such as alcohol. Researchers have developed an ambient-temperature catalytic process for $H_2$ production and simultaneous plastic degradation that not only does not need any extra sacrificial agent but also converts plastic waste to useful organic chemicals. The process, which is known as photoreforming and shown schematically in Fig. 1a [4,5], comprises four essential elements of catalyst, substrate (plastics), water, and light. Under light irradiation, electrons in the valence band (VB) of the catalyst are excited to the conduction band (CB). The excited electrons reduce $H^+$ of water to form $H_2$, while the holes left in VB together with hydroxyl radicals oxidize plastics to other organic products [5,8-10].

A catalyst for simultaneous $H_2$ production and plastic degradation should have several features such as a suitable bandgap to absorb light photons and produce electrons and holes, an appropriate band structure for water reduction and plastic degradation, an appropriate lifetime of electrons and holes, and a stable chemical structure under alkaline conditions. Currently, several groups of catalysts, including sulfides [5,11-14], nitrides [15-18], phosphides [19,20], metal oxides [21,22], and bimetallic metal-organic frameworks (MOFs) [23] are used for this process. While phosphides and sulfides demonstrate the best activity for photoreforming, they usually have low stability compared to oxides [13-16]. Despite the development of such catalysts, particularly in recent years, there is still high demand to explore new families of materials that are highly stable and active for converting plastic waste and producing $H_2$. One possible candidate for this application is high-entropy ceramics which have recently gained attention due to their remarkable stability and promising functional properties including catalytic activity [24-27]. High-entropy ceramics are typically defined as multi-component materials with at least five principal cations and a configurational entropy higher than $1.5R$ ($R$: the gas constant) [28-30]. The high stability of these materials is due to high configurational entropy and accordingly low Gibbs energy, while their unique functional properties are due to their strained lattice, heterogenous valence electron distribution, and cocktail effect [28-30]. The most popular high-entropy ceramics are high-entropy oxides (HEOs), which have been employed in numerous applications such as Li-ion batteries [31,32], Li-S batteries [33], Zn-air batteries [34], dielectric components [35], magnetic components [36], thermal barrier coating [37], and catalysts [38-41]. Despite the high potential of HEOs for catalysis, there have been no reports to date to employ them for photoreforming for simultaneous $H_2$ production and plastic degradation.

In this study, $TiZrHfTaNbO_{11}$, an HEO catalyst, is synthesized and its catalytic activity for simultaneous $H_2$ production and PET plastic degradation is investigated. This HEO is selected because five transition elements of titanium, zirconium, niobium, hafnium, and tantalum have the $d^0$ electronic configuration which is suitable for photocatalysis and their mixture can thus show catalytic properties under light [39]. This first application of HEOs for photoreforming and simultaneous $H_2$ production and plastic conversion reveals that $TiZrHfTaNbO_{11}$ exhibits superior activity compared to the benchmark photocatalyst P25 $TiO_2$, introducing HEOs as a new family of catalysts for photoreforming.

## 2. Materials and methods

The HEO was synthesized by a three-step synthesis method of arc-melting, high-pressure torsion (HPT) severe plastic deformation processing, and high-temperature oxidation, and examined by various characterization methods, as illustrated below.

### 2.1. Reagents

Pure metals of Ti (99.9%, Furuuchi Chemical, Japan), Zr (99.7%, Furuuchi Chemical, Japan), Hf (99.7%, Sigma-Aldrich, USA), Nb (99.9%, Kojundo Chemical, Japan), and Ta (99.9%, Furuuchi Chemical, Japan) were used without any further purification. Pure PET micropowder with particle sizes smaller than 300 μm was purchased from GoodFellow, England, and a PET plastic water bottle with a thickness of 160 μm was purchased from a local market, dried, and cut into pieces with a diameter of 6 mm. NaOH was purchased from Fujifilm, Japan, and a solution of 40% NaOD in $D_2O$ (99 at% of D) was obtained from Sigma-Aldrich, USA. Coumarin as a hydroxyl radical trapper was prepared from Tokyo Chemical Industry Co., LTD, Japan. Isopropanol was purchased from Sigma-Aldrich, USA.

## 2.2. Synthesis of catalyst

The HEO oxide was synthesized in three steps including arc-melting, HPT processing, and oxidation. The illustration of the HPT method is shown in Fig. 1b [42,43], and the sample synthesis procedure is shown in Fig. 1c. Initially, high-purity Ti, Zr, Hf, Nb, and Ta pieces, possessing equal atomic fractions, were arc-melted using a non-consumable tungsten electrode and a water-cooled copper crucible in a high-purity Ar atmosphere to achieve an ingot of high-entropy alloy (HEA) TiZrHfTaNb with 10 g mass. To ensure a homogeneous chemical composition, the resulting HEA ingot was accurately rotated and re-melted seven times. For further enhancement of the homogeneity, the HEA was subjected to HPT. For HPT processing, the HEA ingot was first cut into disc-shaped samples with 10 mm in diameter and 0.8 mm in thickness using an electric-discharge wire cutting. The HPT process was conducted for 10 turns under a pressure of 6 GPa at ambient temperature with a rotation speed of 1 rpm. In the third step, the HPT-processed discs were exposed to air oxidation at 1373 K for 48 h, which resulted in a 22.5% increase in the sample mass, which reasonably corresponds to a composition of $TiZrHfTaNbO_{11}$.

## 2.3. Catalyst characterization

The crystal structure of as-prepared HEO was measured by X-ray diffraction (XRD) using a Cu Kα source. Raman spectra were also recorded with a 325 nm laser as an excitation source. The microstructure and distribution of elements at the micrometer level were examined by scanning electron microscopy (SEM) equipped with energy-dispersive X-ray spectroscopy (EDS). The samples for SEM were dispersed onto the carbon tape and analyzed under 15 kV. The microstructure and the distribution of elements at the nanometer levels were examined by transmission electron microscopy (TEM) and scanning-transmission electron microscopy (STEM) equipped with an EDS detector. For TEM and STEM, the samples were crushed in ethanol and dispersed on copper grids covered with a carbon film. TEM and STEM were conducted in the bright-field (BF) mode, the dark-field (DF) mode, selected area electron diffraction (SAED), and high-resolution (HR) imaging combined with fast Fourier transform (FFT). The specific surface area of catalysts was estimated by $N_2$ adsorption and the Brunauer-Emmett-Teller (BET) approach which was 0.7 $m^2/g$.

Optical properties were first examined by ultraviolet-visible (UV-Vis) diffuse reflectance spectroscopy. Bandgap was calculated by considering the light absorbance using the Kubelka-Munk analysis. To examine the electron-hole recombination, steady-state photoluminescence (PL) was conducted using a 325 nm laser source. The lifetime of excited electrons was examined by a time-resolved photoluminescence decay (PL decay) with a 285 nm laser source. To evaluate the mobility of photoexcited electrons, photocurrent was examined. For photocurrent measurement, fluorine-doped tin oxide glass was coated with a mixture of catalyst and ethanol. Then, to enhance the binding between the catalyst particles and glass substrate, the coated glass was baked in a muffle furnace at 473 K for 1 h. The potentiostatic amperometry mode of photocurrent measurement was performed in $Na_2SO_4$ (0.1 M) electrolyte using the glass as a working electrode, Pt wire as the counter electrode, and Ag/AgCl as a reference electrode. To investigate the generation of oxygen vacancies, electron paramagnetic resonance (EPR) analysis was performed at room temperature using a 9.4688 GHz microwave source.

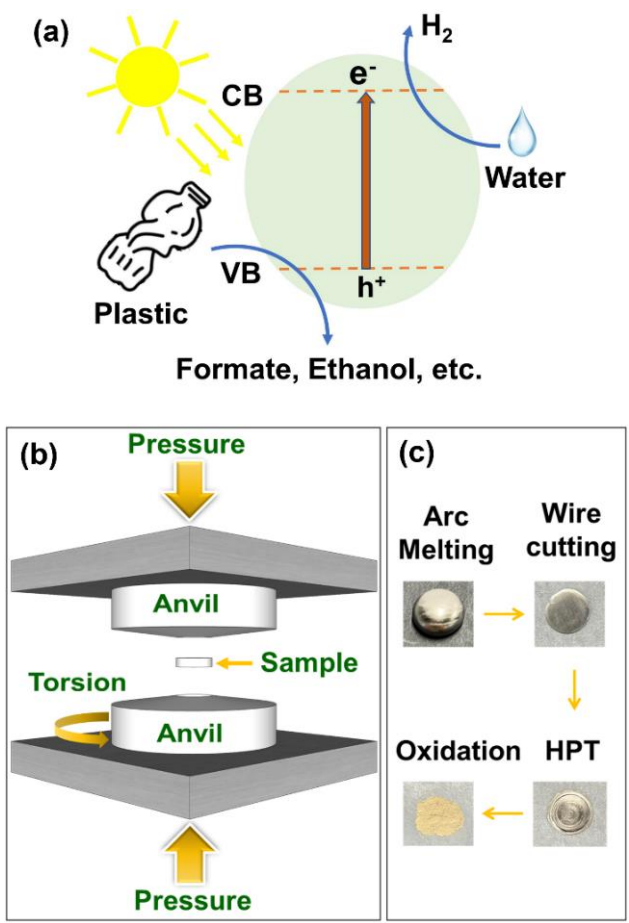

Figure 1. (a) Schematics of photoreforming process, (b) illustration of HPT method, and (c) illustration of this study's steps of synthesis process.

**2.4. Catalytic experiments**

50 mg of PET and 50 mg of the HEO were dispersed in 3 mL of 10 M NaOH mixed with 250 μL of 0.01 M $Pt(NH_3)_4(NO_3)_2$ (to deposit 1 wt% of Pt as a co-catalyst) in a quartz photoreactor tube with a volume of 30 mL. The alkaline NaOH is used to hydrolyze PET to terephthalate and ethylene glycol during photoreforming. A concentration of 10 M of NaOH was selected based on the literature and optimization by the authors [4,11]. The mixture was capped with a rubber septum, air-evacuated by Ar injection for 30 min, and ultrasonicated for 5 min. The samples were then irradiated by a 300 W Xe light. The distance between the light source and the sample was 10 cm and the intensity of the light source at this distance was 18 kW/m$^2$. The temperature of the photoreactor was kept at 298 K by using a water bath, while stirred with a rate of 800 rpm. H$_2$ generation was quantified by gas chromatography using a thermal conductivity detector with Ar being the carrier gas. For comparison purposes, P25 TiO$_2$ as a reference catalyst was also examined under similar reaction conditions.

### 2.4. Post-catalysis characterizations

The stability of the catalyst after the catalytic test was examined by conducting XRD and Raman analyses and comparing the results with the initial HEO. The reusability of the HEO was assessed through three consecutive photocatalytic cycles. After each cycle, the HEO was collected, thoroughly rinsed with deionized water until the pH was neutral, dried at 333 K, and then reused in the photoreforming process.

For the evaluation of the plastic degradation products and radicals, the liquid phase was examined by nuclear magnetic resonance (NMR) and fluorescence spectroscopy. The concentration of hydroxyl (•OH) radical in the liquid phase after the catalytic process was measured by adding 0.4 mg of coumarin in the form of 1 mM solution. Since coumarin reacts with •OH and produces 7-hydroxycoumarin, a spectrofluorometer with an excitation wavelength of 332 nm was employed to record the fluorescence signal of the generated 7-hydroxycoumarin at 450 nm as an indicator of the •OH concentration. The quenching experiment with isopropanol (0.02 mmol/L) was conducted to investigate the role of •OH in the photoreforming reactions. $^1$H NMR spectra were collected by a Bruker spectrometer using 10 M NaOD in D$_2$O to identify the degradation products of PET plastic during catalytic degradation. The noesygppr1d process program was used for the $^1$H NMR analysis with water suppression. Maleic acid with a concentration of 50 mg mL$^{-1}$ in D$_2$O was utilized for quantitative analysis. The quantity of oxidized products ($m_{product}$) after the photoreforming reaction was calculated by Eq. 1.

$$m_{\text{product}} = \frac{I_{\text{product}}}{I_{\text{maleic acid}}} \times \frac{N_{\text{maleic acid}}}{N_{\text{product}}} \times \frac{M_{\text{product}}}{M_{\text{maleic acid}}} \times m_{\text{maleic acid}} \qquad (1)$$

with $m$, $I$, $N$ and $M$ are the mass, integral of the NMR peak, number of protons corresponding to the peak and molar mass, respectively.

The plastic pieces were also examined by different methods. An analytical balance was used to measure the degraded plastic mass during catalysis. Remained plastic pieces were sputter-coated with a thin layer of gold and analyzed by SEM. Attenuated total reflectance-Fourier transform infrared spectroscopy (ATR-FTIR) within the range of 600-4000 cm$^{-1}$ was utilized to analyze the possible changes in the structure of plastic during catalytic degradation.

### 3. Results
### 3.1. Catalyst characterization

Crystal structure evolution during each synthesis step was examined by XRD profiles, as shown in Fig. 2a. The ingot after arc melting possesses a body-centered cubic (BCC) phase with a space group of *Im-3m* (229) and a lattice parameter of $a = 3.41\pm0.21$ Å. This is consistent with previous reports on similar HEA [44,45]. The same BCC phase of HEA is obtained after the second synthesizing step of HPT, but the main difference is the significant peak broadening due to the crystallite size reduction and defect generation, as reported in a wide range of HPT-processed materials [42,43]. Oxidized samples contain two phases with monoclinic and orthorhombic structures with almost equal mass fractions. The monoclinic phase has a *A2/m* space group with the cell parameters of $a = 11.87\pm0.07$ Å, $b = 3.82\pm0.15$ Å, $c = 20.55\pm2.05$ Å, $\alpha = \gamma = 90º$, and $\beta = 119.93º\pm4.80º$ [46]. The orthorhombic phase has an *Ima*2 space group with the cell parameters of $a = 45.97\pm1.84$ Å, $b = 4.62\pm0.23$ Å, $c = 5.12\pm0.26$ Å, and $\alpha = \beta = \gamma = 90º$ [47]. Raman spectra taken from different positions of HEO confirm similar spectra, as shown in Fig. 2b, but peak intensities follow two distinguishable patterns which should be due to the presence of dual phases.

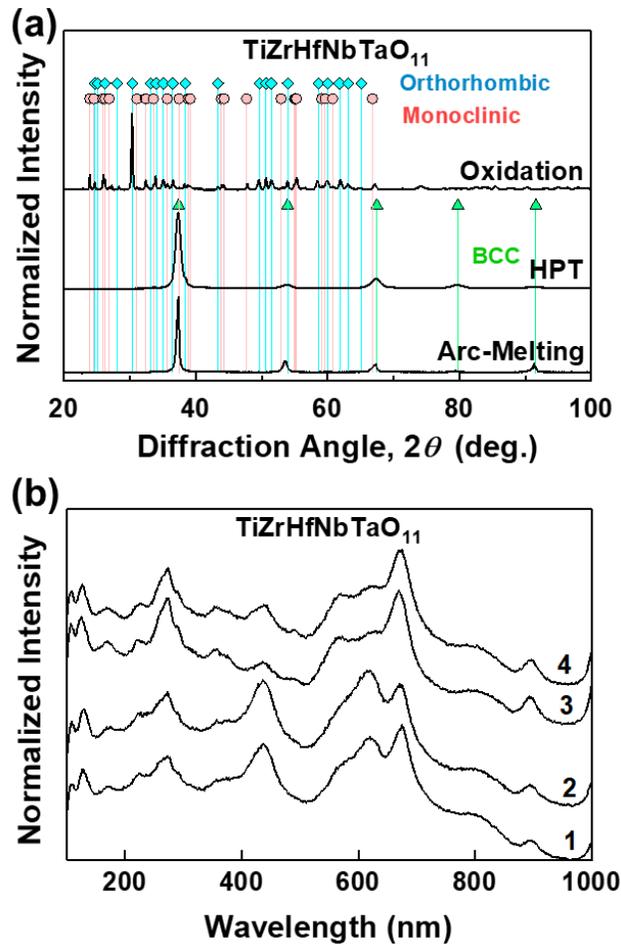

Figure 2. Dual-phase structure of high-entropy oxide examined by (a) XRD and (b) Raman spectroscopy taken at four different positions.

The microstructure of the HEO sample, examined by SEM micrographs at different magnifications is shown in Fig. 3. SEM measurement indicates that the average particle size is 1.9 µm, while the examination of microstructure in a higher magnification indicates that there are numerous crystals in particles with an average size of 310 nm. These rather large particle sizes are due to the synthesis method which uses high pressure and high temperature, leading to the consolidation of HEOs [30].

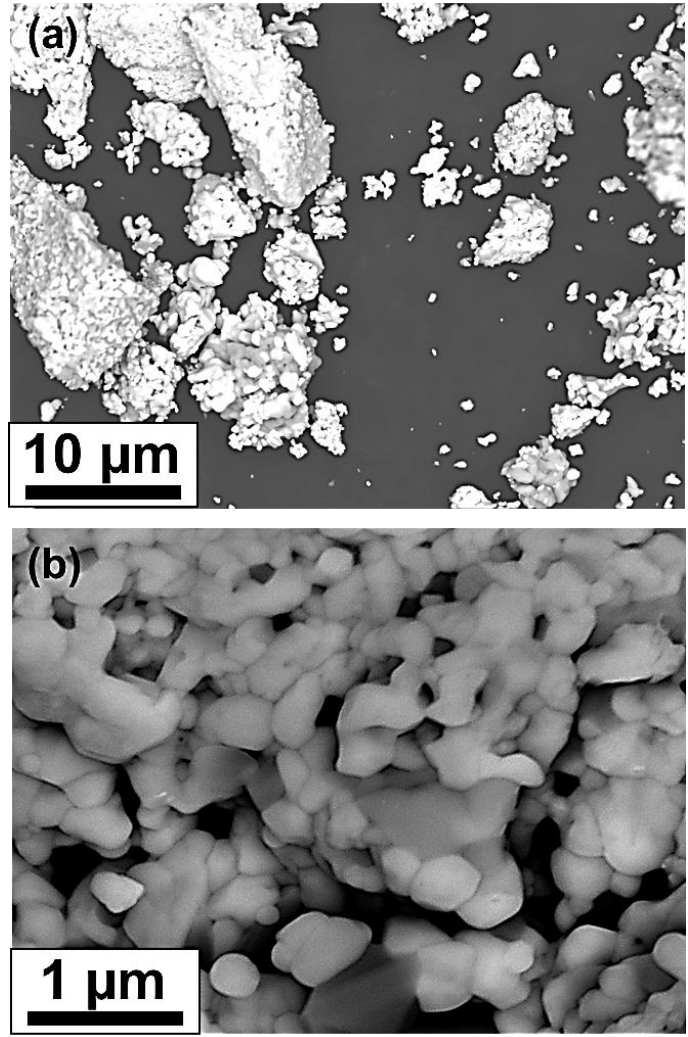

Figure 3. Morphology of high-entropy oxide examined by SEM at different magnifications.

High-entropy materials usually possess a uniform distribution of elements [30-37]. To confirm the uniformity of elements in the HEO, SEM-EDS, and STEM-EDS were performed, as shown in Fig. 4. Despite the dual-phase feature of the HEO, the distribution of five elements is reasonably homogeneous. SEM-EDS analysis also suggests that the general composition of the HEO should be TiZrHfTaNbO$_{11}$ which is consistent with the mass analysis before and after arc melting as well as with mass analysis before and after oxidation. Moreover, the valence of cations in this system suggests a composition of TiZrHfTaNbO$_{11}$ which is consistent with SEM-EDS and mass analyses [38].

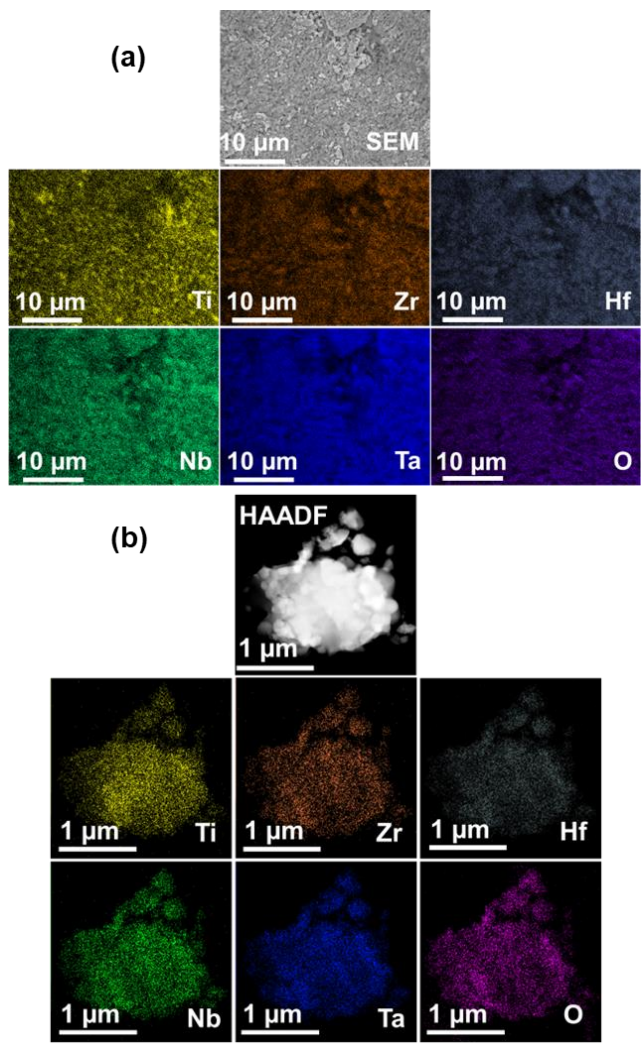

Figure 4. Uniform distribution of elements in high-entropy oxide. (a) SEM-EDS and (b) STEM-EDS analyses.

Nanostructural examinations are shown in Fig. 5 using TEM in (a) BF, (b) SAED, and (c-f) HR modes. The BF image and ring pattern of SAED verify the presence of numerous nanocrystals with random crystallographic orientations. The lattice image in Fig. 5c exhibits significant lattice distortion with numerous dislocation defects within the crystal structure. The presence of dislocations can be considered a positive feature in this work because previous studies indicated that dislocations could amplify light absorption and improve catalytic activity [48]. The lattice images in Fig. 5d-f support the co-existence of monoclinic and orthorhombic phases and a significant proportion of interphase boundaries. The interphases can serve as charge heterojunctions to facilitate the separation and mobility of charge carriers to increase catalytic activity [49]. Here, it should be noted that EPR spectroscopy did not provide a clear piece of evidence for oxygen vacancies in this HEO, although vacancies are generally expected to exist in high-entropy materials [28-30].

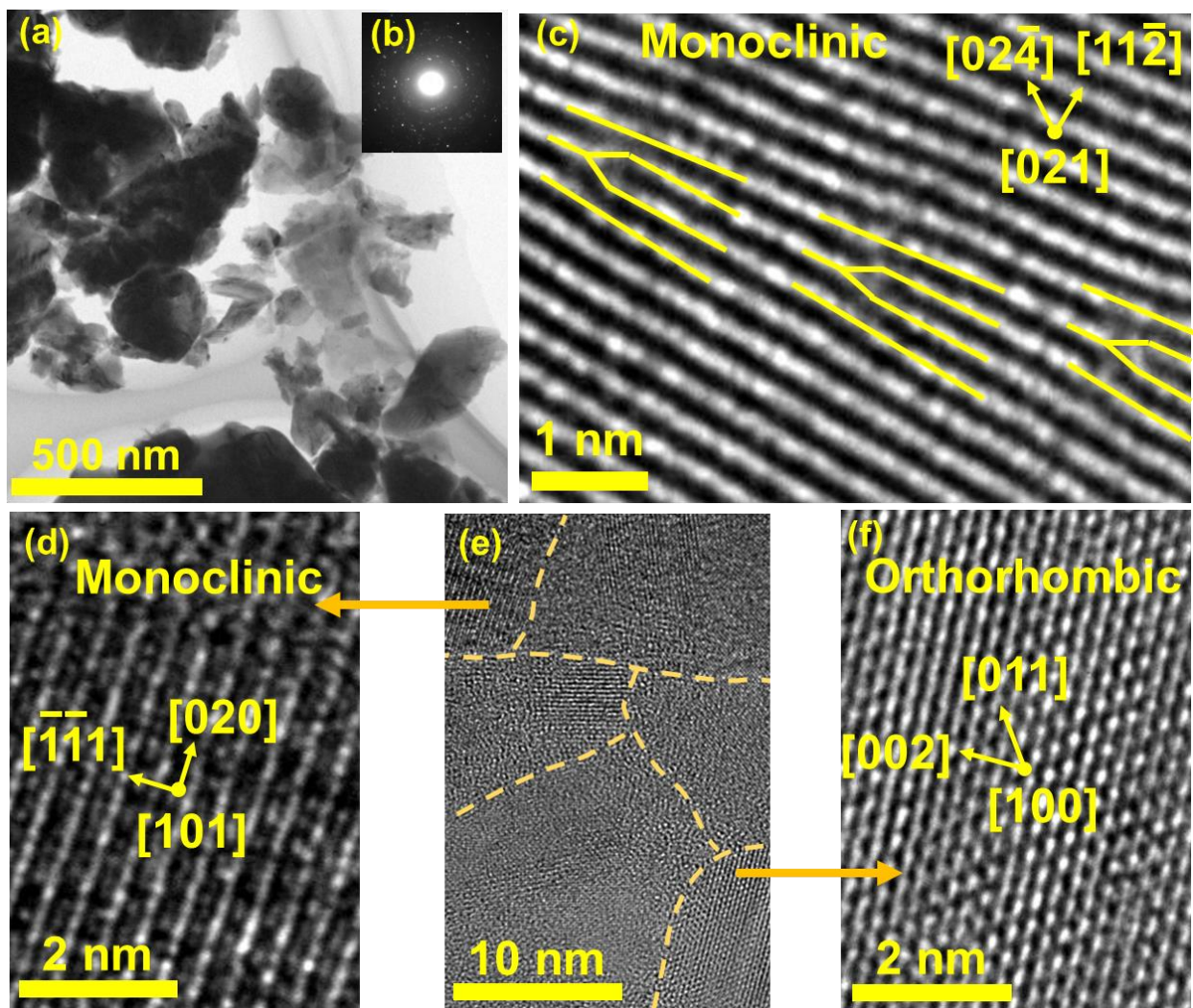

Figure 5. Formation of dual-phase high-entropy oxide with high density of dislocations and interfaces. TEM (a) BF image, (b) SAED pattern, and (c-f) HR images at different magnifications.

Fig. 6a presents the UV-Vis spectrum of HEO with an obvious absorbance edge at 430 nm and some light absorbance in the visible light region. For comparison, the UV-Vis spectrum of P25 $TiO_2$, as a benchmark catalyst, is included in Fig. 6a. The absorbance edge of HEO, which is similar to P25, is related to the bandgap of the material, while the defects might be considered as the reason for the higher visible light absorbance of HEO compared to P25. Based on the Kubelka-Munk analysis, the bandgap is estimated at 3.2 eV, as shown in Fig. 6b. This is reasonably similar to the $TiO_2$ bandgap and narrower than that of other binary oxides in the Ti-Zr-Hf-Nb-Ta-O system which were reported from 3.2 eV to 5.7 eV [50].

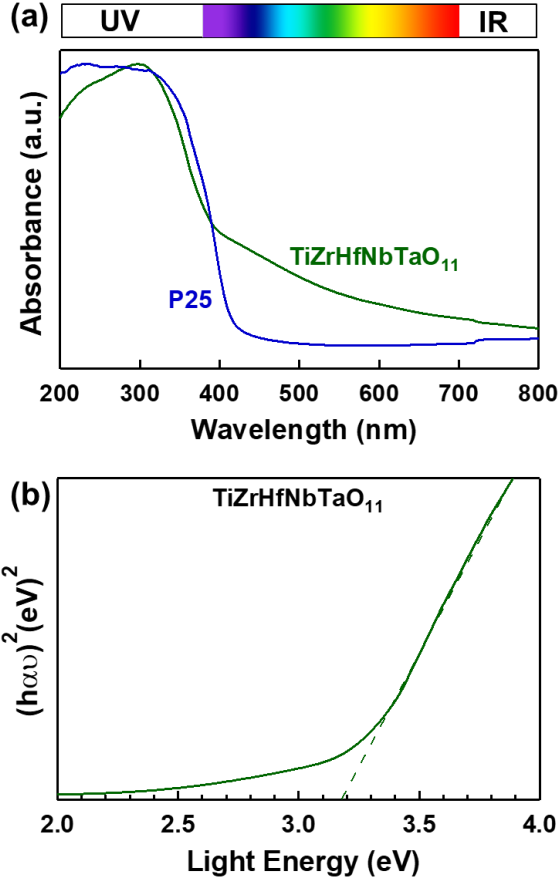

Figure 6. Light absorbance of high-entropy oxide in both UV and visible light regions. (a) UV-Vis light absorbance spectrum of high-entropy oxide compared to P25 $TiO_2$, and (b) Kubelka-Munk plot for bandgap calculation ($\alpha$: light absorption, h: Planck's constant, $\nu$: light frequency).

Charge-carrier dynamics and oxygen vacancies were examined by various techniques including steady-state PL, time-resolved PL decay, and photocurrent, as shown in Fig. 7. The steady-state PL spectra show a PL peak of HEO at 552 nm, which should be due to radiative recombination of electrons and holes on defects. The PL peak intensity of the HEO is comparatively low compared to the PL peak intensity of P25 (Fig. 7a). This suggests that the current HEO can serve as a good catalyst with low electron-hole recombination. PL decay spectrum in Fig. 7b follows an exponential equation [51].

$$I(t) = A_1 \exp\left(-\frac{t}{\tau_1}\right) + A_2 \exp\left(-\frac{t}{\tau_2}\right) \tag{2}$$

In this equation, $I(t)$ represents the PL decay intensity at time $t$, $A_1$ and $A_2$ are the amplitudes of the first and second exponential functions, respectively, and $\tau_1$ and $\tau_2$ correspond to the fast and slow decay time. The analysis of the PL decay yields values of $A_1 = 42.3$, $A_2 = 57.7$, $\tau_1 = 1.5$ ns, and $\tau_2 = 10.4$ ns. Subsequently, the average electron lifetime of the HEO, $\tau_{ave}$ was calculated as 9.5 ns based on the following equation [51].

$$\tau_{ave} = \frac{A_1\tau_1^2 + A_2\tau_2^2}{A_1\tau_1 + A_2\tau_2} \tag{3}$$

This lifetime is considered rather long for photocatalysts. For example, the previously reported lifetime values of the most common $TiO_2$ catalysts are 0.93 ns for anatase [52], 0.03 ns for rutile [53], and 1.55 ns [54] or 2.41 ns [55] for P25. The longer charge carrier lifetime of the current HEO indicates its high potential for superior photoreforming. Photocurrent measurements in Fig. 7c confirm that the HEO successfully generates photo-induced electrons. For comparison, Fig. 7d shows the photocurrent generation of P25. Here, it should be noted that P25 has nanopowders that make a large interface area with FTO glass for electron transfer, while the HEO has micropowders with a small interface area with FTO. Therefore, a direct comparison of photocurrent intensities does not provide much information about the behavior of the two materials. However, the differences in the shape of photocurrent spectra give important information. For P25, spike peaks are visible at the start of illumination followed by a significant decrease in photocurrent, indicating that the generation of electrons and holes is followed by quick recombination. For the HEO, the spike peaks are absent except for the first illumination cycle, suggesting a better electron-hole separation than P25.

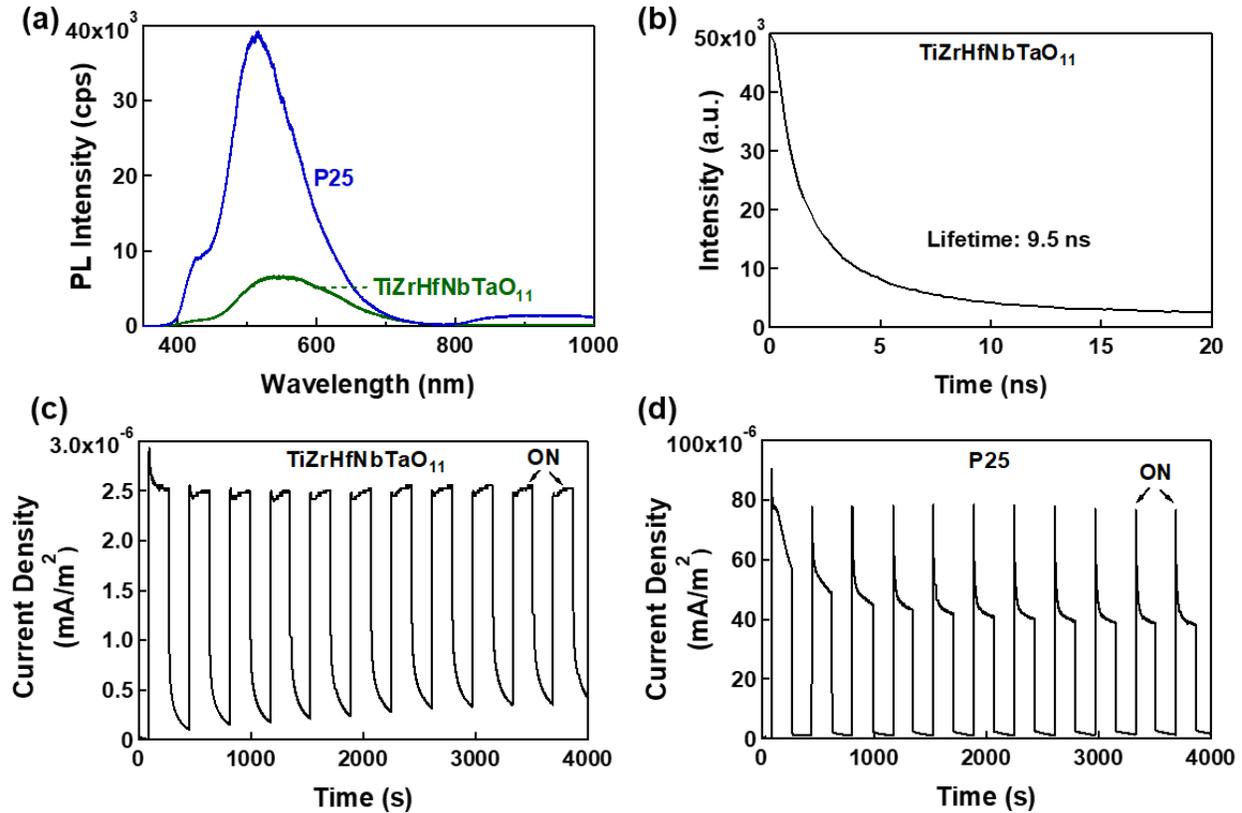

Figure 7. Reasonable charge carrier dynamics of high-entropy oxide. (a) Steady-state PL emission of HEO and P25, (b) time-resolved PL decay of HEO, (c) photocurrent generation of HEO, and (d) photocurrent generation of P25.

## 3.2. Photoreforming of PET microplastics

The catalytic activity of HEO in comparison with P25 for $H_2$ production from PET microplastic photoreforming under light with and without platinum co-catalyst is presented in Fig. 8. No $H_2$ is detected for blank tests without a catalyst addition and with irradiation (red curve) or with catalyst addition under the dark conditions (data at time zero). The amount of produced $H_2$ increases as the photo-exposure time increases, indicating a direct correlation between photo-exposure time and $H_2$ production yield. During the first 30 min, there is a higher rate of $H_2$ production, but the rate of $H_2$ production subsequently reduces and follows a linear curve. The faster rate at the first 30 min can be attributed to the abundance of initial active sites on both catalysts and PET plastic surfaces [4,5]. Fig. 8a shows that the addition of platinum co-catalyst enhances the $H_2$ production in both P25 and HEO systems which signifies the important role of co-catalyst in providing active sites for the photoreforming process. Since the comparisons in Fig. 8a can be influenced by a large difference between the surface area of the two catalysts (HEO and P25), additional comparisons were made by $TiO_2$ polymorphs (rutile, anatase and brookite) and ZnO which are well-known photocatalysts for hydrogen production. To bring the surface area of these $TiO_2$- and ZnO-based catalysts closer to the HEO, HPT processing was applied to them, as attempted in a few earlier studies [56-59]. Although HPT processing can significantly enhance the photocatalytic activity by the introduction of oxygen vacancies and phase transformations (e.g. anatase to columbite in $TiO_2$ [56,57] and wurtzite to rocksalt in ZnO [59]), it is still interesting to compare such catalysts with the HEO for photoreforming. As shown in Fig. 8b, anatase-columbite $TiO_2$ processed by HPT for 3 turns (surface area: 0.7 $m^2/g$ [57]) and black brookite $TiO_2$ processed by HPT at 473 K for 1 turn (surface are: 0.2 $m^2/g$ [58]) exhibits a better activity than HEO perhaps due to easy charge separation, slow decay and moderate electron trap energy in black brookite [60], and high water adsorption energy and better surface activity for proton formation in columbite [61]. However, the activity of the HEO (surface area: 0.7 $m^2/g$) is better than rutile $TiO_2$ synthesized by HPT for 3 turns followed by annealing at 1073 K and ZnO processed by HPT for 4 turns (surface area: 1.3 $m^2/g$ [59]).

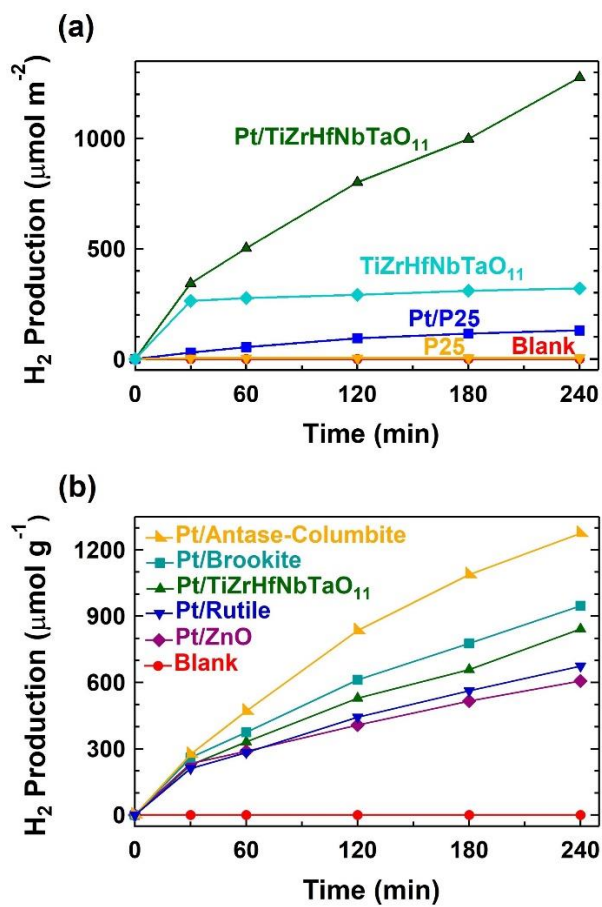

Figure 8. High activity of high-entropy oxide for hydrogen production from photoreforming of PET microplastics. (a) $H_2$ production versus irradiation time for high-entropy oxide and P25 $TiO_2$ catalysts with and without co-catalyst including blank test. (b) $H_2$ production versus irradiation time for high-entropy oxide with co-catalyst addition compared with rutile $TiO_2$ synthesized by HPT for 3 turns followed by annealing at 1073, anatase-columbite $TiO_2$ processed by HPT for 3 turns, black brookite $TiO_2$ processed by HPT at 473 K for 1 turn and ZnO processed by HPT for 4 turns.

In the photoreforming process, not only green $H_2$ is generated from water through the reduction half-reaction but also other useful products are produced through plastic oxidation. To identify the organic oxidation products from PET, $^1$H NMR analysis was performed in 10 M NaOD in $D_2O$ with the same sample concentration as that of the catalytic test. The $^1$H NMR spectra are shown in Fig. 9 and the quantitative data are presented in Table 1, indicating PET is degraded into different products after 4 h of light irradiation. It can be observed that PET undergoes hydrolysis to ethylene glycol and terephthalate. Terephthalate is quite stable and is recovered to produce PET plastic again, while ethylene glycol is subsequently photo-converted into formic acid. The PET plastic degradation pathway is illustrated in Fig. 10. Comparable oxidation products of PET have been documented in several previous studies [4,62,63]; however, variations in the degradation products have been also observed, likely due to differences in experimental conditions such as the choice of catalyst, alkaline concentration, irradiation duration, pre-treatment method or the quantity of plastic used. For example, in a previous study, after 8 days of photoreforming, carbonized polymer dots/g-$C_3N_4$ converts PET to 110 µmol glycolaldehyde, 383 µmol glycolic acid, 139 µmol formic acid, 128 µmol ethanol, 43 µmol acetaldehyde, and 554 µmol acetic acid [16]. In another study, carbon nitride - carbon nanotubes - NiMo hybrid converted PET to glyoxal and glycolic acid [3], while $MoS_2/Cd_xZn_{1-x}S$ led to the formation of formic acid, glycolic acid and methylglyoxal from PET [64].

Quantitative data gathered for hydrogen production and plastic degradation can be used to estimate the degradation rate, charge balance and carbon balance. During 4 h of experiments using a solution of 10 M NaOH, the hydrolysis process produces 0.17 mmol m$^{-2}$ terephthalate and 1.43 mmol m$^{-2}$ ethylene glycol. Subsequently, 16% of ethylene glycol is oxidized to produce 0.54 mmol m$^{-2}$ formic acid. The electrons utilized in hydrogen generation (2 electrons per $H_2$) after 4 h of photoreforming are estimated at approximately 2.4 mmol m$^{-2}$, closely matching the hole transfer number (4.3 mmol m$^{-2}$) calculated from the oxidation of ethylene glycol to formic acid (8 holes per HCOOH), resulting in a charge balance of 55%. The carbon balance achieved by comparing the mass of the initial plastic and the mass of degradation products is at least 32%, possibly underestimated due to the remaining undegraded plastics as well as due to not detecting certain oxidized products by NMR. It should be noted that the hydrolysis products were too small to be detected by NMR when NaOH concentration was low and pH was below 14. Even using a 5 M NaOH solution did not lead to measurable hydrolysis products within the detection limes of NMR. These results indicate that a highly alkaline solution is essential for the photoreforming of PET, which is somehow a negative aspect of this process. Future studies should incorporate more sensitive analyses, such as high-performance liquid chromatography (HPLC), to detect a wider range of products.

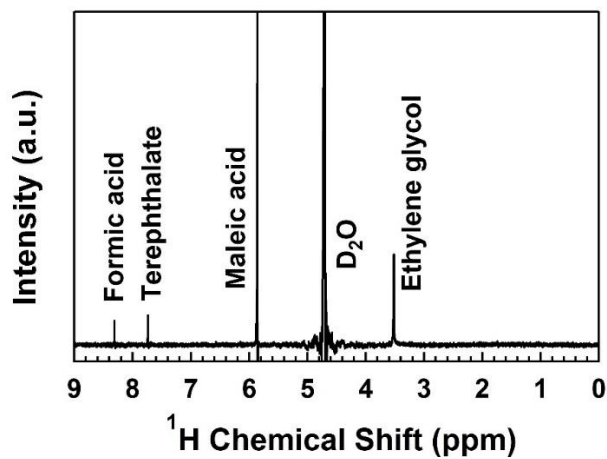

Figure 9. Formation of various valuable chemicals from photoreforming of PET microplastics using high-entropy oxide. $^1$H NMR spectra of PET after 4 h light irradiation in the presence of high-entropy oxide using 10 M NaOD in $D_2O$ and maleic acid as internal standard.

Table 1. Quantitative PET photoreforming products after 4 h light irradiation in the presence of high-entropy oxide, 10 M NaOD in $D_2O$ and maleic acid.

| Products | Amount (mmol) | Amount (mmol/m$^2$) |
|---|---|---|
| $H_2$ | 0.042 | 1.20 |
| Terephthalate | 0.006 | 0.17 |
| Ethylene glycol | 0.050 | 1.43 |
| Formic acid | 0.019 | 0.54 |

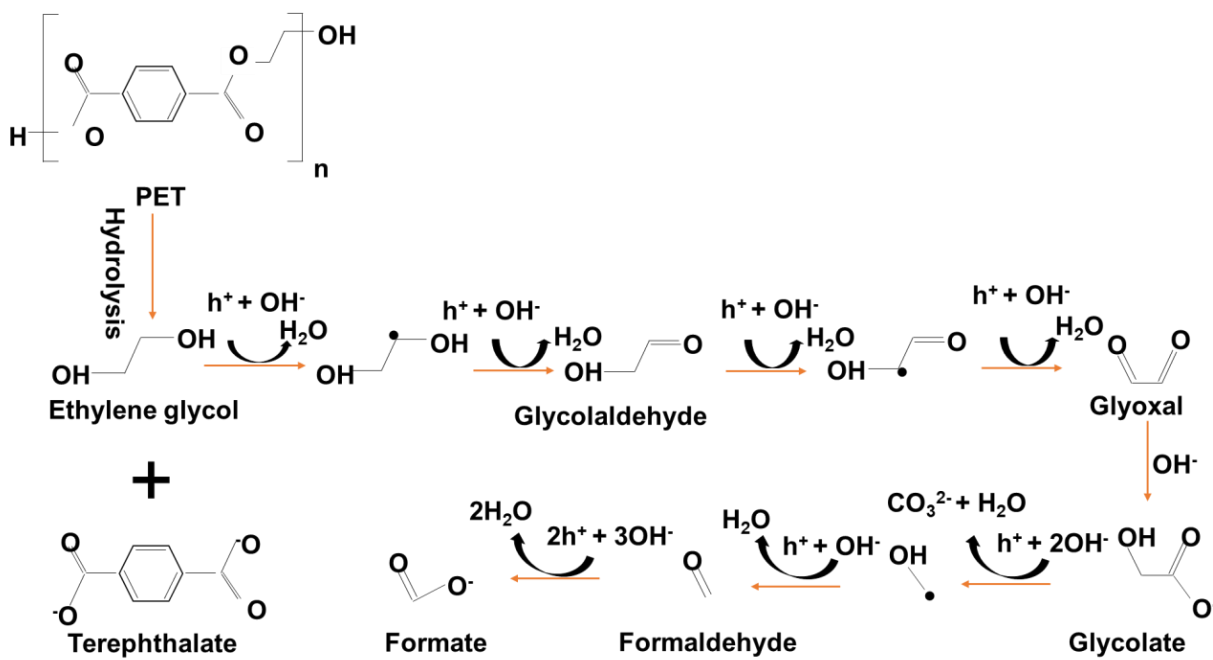

Figure 10. Degradation pathways of PET microplastics using high-entropy oxide.

In the photoreforming process, the generated reactive radicals, particularly •OH radicals, contribute to the oxidation transformations and control the catalyst activities [4,5]. Therefore, it is important to assess the generation of radicals during the process. The quantification of •OH radical generation was conducted by employing coumarin as a probe radical trap reagent, and examination of the formation of coumarin-OH adduct (7-hydroxycoumarin) by fluorescence spectroscopy [64]. Fig. 11a presents a comparison of the production of 7-hydroxycoumarin resulting from the use of the HEO compared with P25 $TiO_2$. The data reveal a notably higher production of 7-hydroxycoumarin from the HEO, signifying that it exhibits a superior rate of •OH radical formation, even in comparison with the benchmark catalyst P25 $TiO_2$. To further investigate the role of •OH radicals on the photoreforming process, the quenching experiment for •OH radical scavenger was conducted. As shown in Fig. 11b, the addition of isopropanol leads to a decrease in $H_2$ production from PET plastic photoreforming. Therefore, •OH radical is one of the important factors in the photoreforming process.

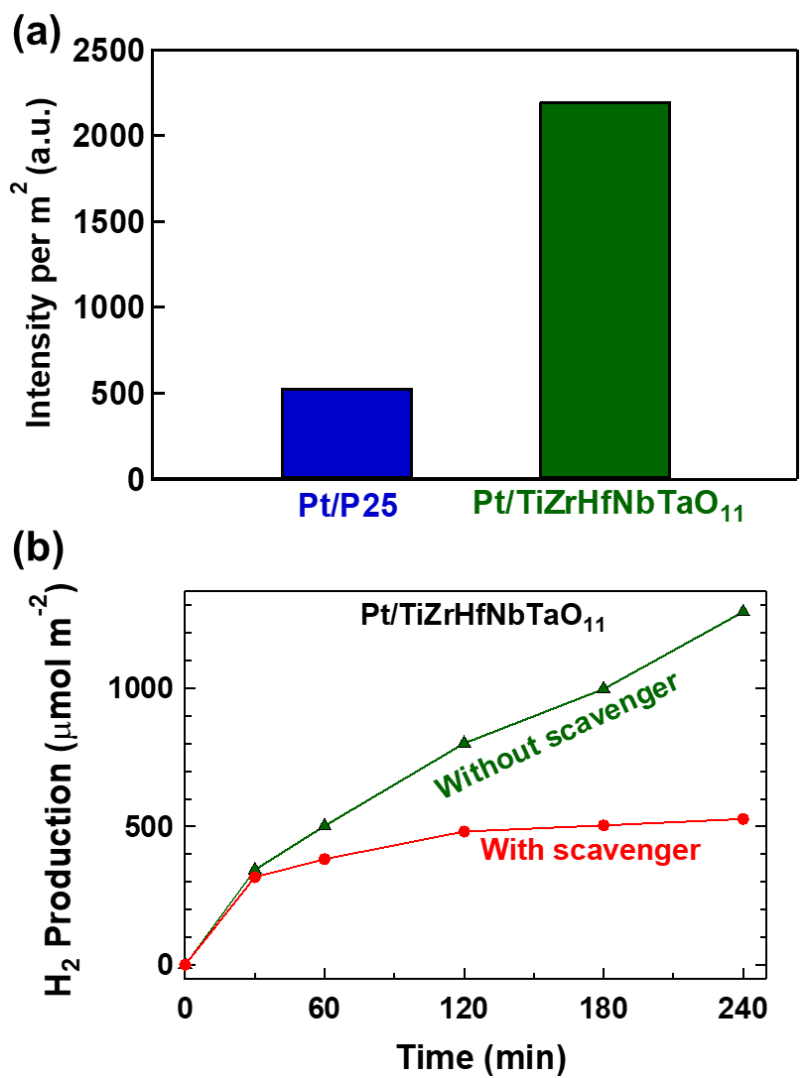

Figure 11. High •OH radical formation from photoreforming using high-entropy oxide. (a) Fluorescent light intensity corresponding to reaction of •OH radicals and coumarin and formation of 7-hydroxycoumarin using high-entropy oxide and P25 TiO$_2$ catalysts after 20 min light irradiation and (b) hydrogen production versus irradiation time during photoreforming process with and without addition of isopropanol (•OH radical scavenger) for Pt/TiZrHfNbTaO$_{11}$.

### 3.3. Photoreforming of PET plastic bottle

In this study, the photoreforming of a PET bottle was also examined. Since commercial plastic bottles often contain additional fillers, antioxidants, or linkers, their degradation behavior can be different from pure PET micropowders. Fig. 12 shows catalytic H$_2$ production versus light irradiation time for the photoreforming of PET bottle and micropowders. After 4 h of irradiation, photoreforming of PET bottle produces 343 μmol m$^{-2}$ of H$_2$, while the generated H$_2$ amount is as high as 1276 μmol m$^{-2}$ for photoreforming of PET micropowders. A four-time higher activity of PET micropowder compared to the PET bottle can be due to its higher purity as well as larger surface area. Photoreforming of the PET bottle even with a lower rate compared to pure microplastics is still of significance because plastic bottles are responsible for a large environmental contamination [1,2]. Here, it should be noted that although plastic waste cannot enhance the light absorbance by the catalyst for photoreforming, the charge carrier transition from catalyst to plastic can positively enhance the lifetime of charge carriers [65].

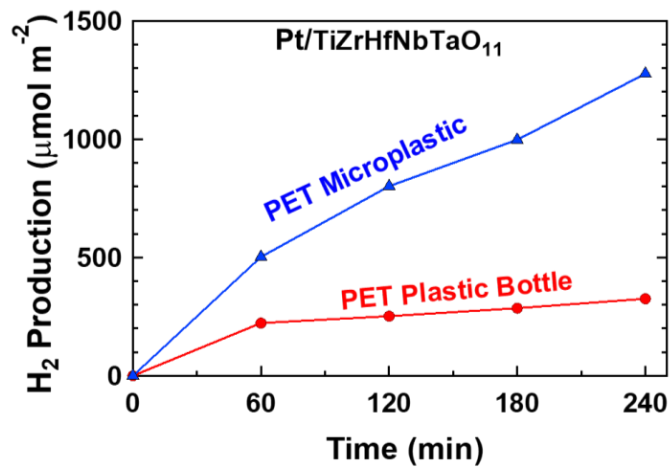

Figure 12. Slower photoreforming rate of PET plastic bottle compared to micropwders using high-entropy oxide. Hydrogen production versus irradiation time for PET bottle and microplastics using high-entropy oxide and P25 $TiO_2$ catalysts.

Fig. 13 shows the change in (a) light transmission, (b) mass loss, and (c, d) surface morphology of PET plastic, examined by SEM, before and after the photoreforming process. The mass loss was calculated as

$$\text{Mass loss (\%)} = \frac{m_0 - m_t}{m_0} \times 100 \tag{4}$$

where, $m_0$ and $m_t$ represent the initial and final masses of PET pieces after photodegradation. The light transmittance of PET plastic slightly decreases over time, as shown in Fig. 13a. In other words, the surface of the initial PET sample is clean, and its light transmittance is high; however, as the degradation progresses, it becomes slightly opaque due to the development of surface roughness and micro-holes. The mass loss, shown in Fig. 13b versus irradiation time, indicates a significant increase in plastic mass with irradiation time. After 16 h of irradiation, 100% of the plastic pieces are estimated to be photodegraded. Fig. 13c presents the spotlessness of the initial PET surface; however, an increase in surface roughness and formation of deep cavities and holes are observed after irradiation (Fig. 13d). This indicates that PET erosion occurs as a result of the photodegradation process which leads to transparency and mass losses as already shown in Fig. 13a and 13b. The observation of the photoreactor during irradiation also confirmed the gradual degradation of plastics and loss of transparency due to the formation of terephthalate monomers over time. These results indicate that photoreforming and mass loss of PET bottles proceed within a rather long time, but such a reaction can be accelerated by the mechanical treatment of bottles to form fine powders [1,2].

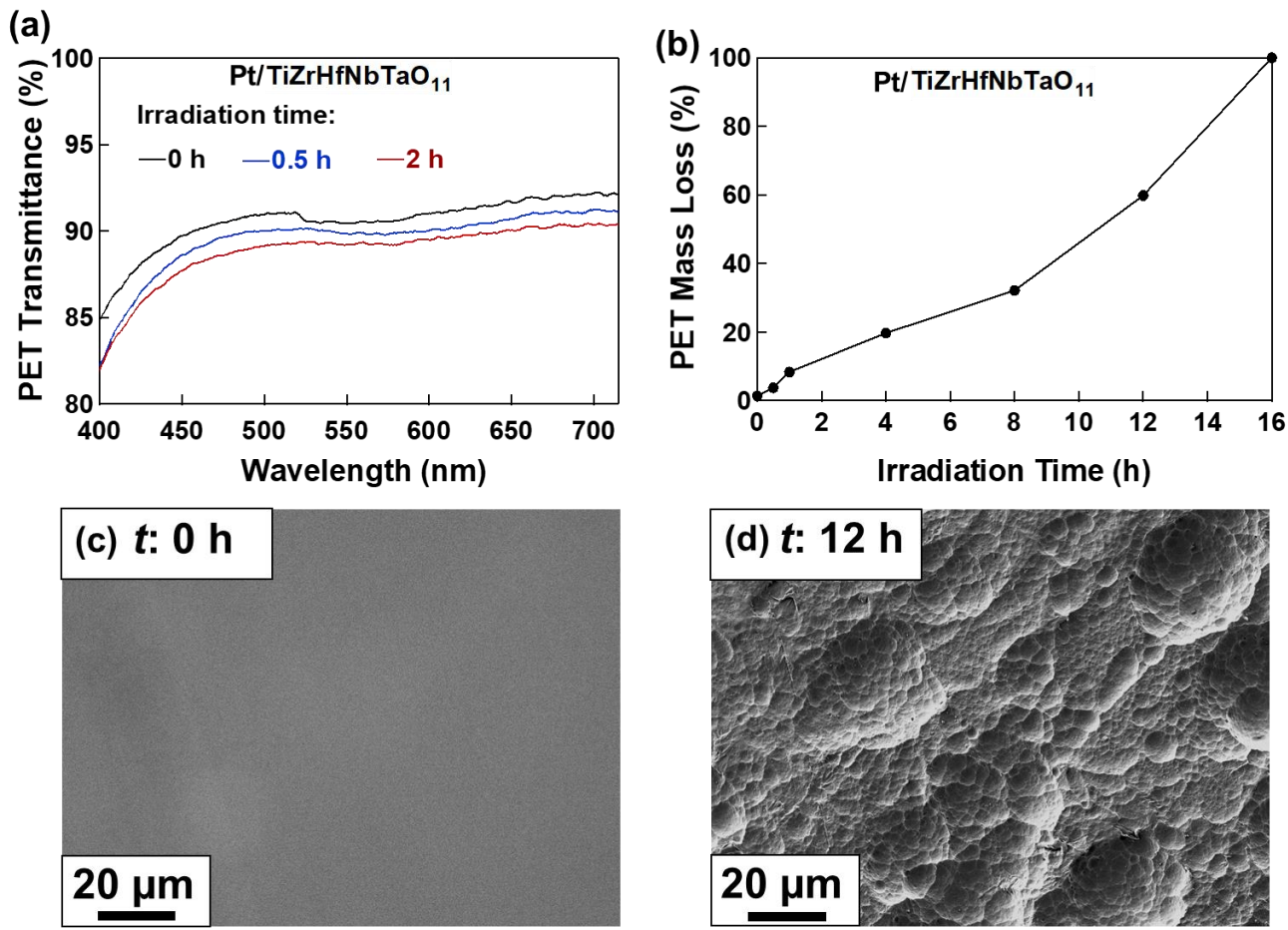

Figure 13. Loss of transparency and mass of PET plastics during photoreforming using high-entropy oxide catalyst. (a) UV-Vis transmission spectra of PET plastic bottle before and after photoreforming for 0.5 and 2 h, (b) mass loss of PET plastic bottle versus irradiation time, and (c, d) surface morphology of PET plastic bottle examined by SEM (c) before and (d) after photoreforming for 12 h.

ATR-FTIR spectroscopy was utilized to investigate the deterioration of PET pieces through oxidative degradation by the catalytic process. ATR-FTIR spectra were compared before and after the process for 12 h and shown in Fig. 14. As illustrated in Fig. 14a for the initial PET plastic bottle, the main identified peaks are C-H bending ethyl (731 cm$^{-1}$), C-O ester (1050-1100 cm$^{-1}$), C-C phenyl ring (1408 cm$^{-1}$), C=O ester (1725 cm$^{-1}$), H-C=O group (2907 cm$^{-1}$), and C-H ethyl (2970 cm$^{-1}$) stretching vibration modes [66]. A comparison between the ATR-FTIR spectra before and after photoreforming, as shown in Fig. 14b, indicates several points. (i) Peak intensities mostly reduce with photoreforming due to the photodegradation-induced elimination of ketones and aromatic characters in PET. (ii) The peak shift at vibration mode of 845 cm$^{-1}$ affirms the deterioration of aromatic rings [66]. (iii) The peak shift at the vibration mode of 1725 cm$^{-1}$, which corresponds to the C=O stretching mode, indicates the formation of a terminal carboxylic acid group in the molecule. (vi) Several characteristic vibration modes at 1245, 1125, 1340, 975, and 845 cm$^{-1}$, corresponding to carbonyl, benzene, ether, and methylene groups, exhibit peak shifts due to the trans conformer. (v) IR bands at 1370 and 1040 cm$^{-1}$, which are attributed to the ethylene glycol unit vibrations, shift due to the gauche conformer. (vi) The IR band at 975 cm$^{-1}$, resulting from the asymmetric stretching of the trans-oxy-ethylene (O-CH$_2$) group, shifts due to morphological changes [65]. To quantify ATR-FTIR data, the carbonyl index (*CI*) and vinyl index (*VI*) were determined as follows [8].

$$CI = \frac{A_{1740} - A_{1835}}{0.08 \times t} \tag{5}$$

$$VI = \frac{A_{909}}{A_{2020}} \tag{6}$$

where, $A_{1740}$ and $A_{1835}$ represent the absorption intensities of the stretching vibration of the carbonyl group at 1740 cm$^{-1}$ and 1835 cm$^{-1}$, respectively, *t* is the sample thickness (0.16 mm), $A_{909}$ and $A_{2020}$ denote the absorption intensities at 909 cm$^{-1}$ and 2020 cm$^{-1}$, respectively. The calculated *CI* for the initial PET bottle and after 12-hour photodegradation are 7.36 and 3.95 as shown in Table 2. Concurrently, the *VI* exhibits a minor change from 0.98 to 0.99 after 12 hours of light irradiation. These variations in *CI* and *VI* are indicative of the photodegradation process occurring in PET plastic under light exposure [8].

### 3.4. Catalyst stability and reusability

Characterization of the HEO catalyst after H$_2$ generation and PET degradation was performed by XRD and Raman spectroscopy, as shown in Fig. 15a and 15b, respectively. Data show that the crystal structure of the oxide is stable under light irradiation and high alkaline conditions during the catalytic test. Both monoclinic and orthorhombic phases remain unchanged after catalytic reactions. The stability of current HEO is consistent with previous studies which reported the remarkable stability of HEOs for numerous applications including Li-ion batteries [31,32], Li-S batteries [33], Zn-air batteries [34], dielectric components [35], magnetic components [36], thermal barrier coating [37], and catalysts [38-41]. The reusability of the HEO catalyst in three repeated cycles is presented in Fig. 15c. The HEO catalyst exhibits good reusability because there is no significant decrease in hydrogen production between the three cycles. The high stability and reusability of HEOs are attributed to their high entropy, leading to a low Gibbs energy that promotes enduring performance through various environmental conditions and operational demands [28-30].

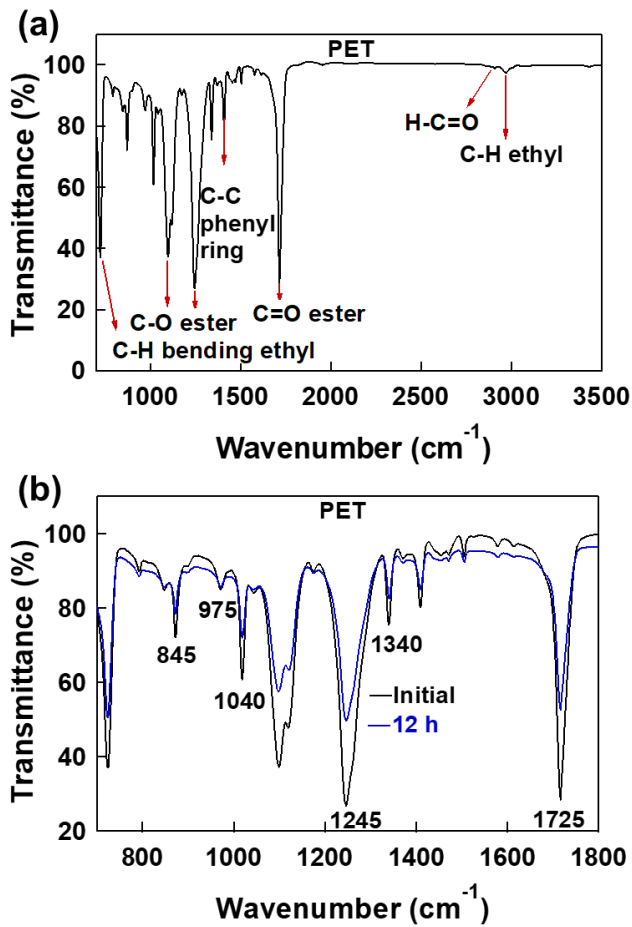

Figure 14. Reduction of the intensity of ATR-FTIR peaks and their shift after photoreforming of PET bottle using high-entropy oxide. ATR-FTIR spectra for (a) initial PET bottle and (b) PET bottle before and after catalysis.

Table 2. Decrease in the carbonyl index (*CI*) and a minor increase in the vinyl index (*VI*) after photodegradation of PET bottle using high-entropy oxide catalyst. ATR-FTIR absorbance intensities at wavenumbers of 1740, 1853, 909, and 2020 cm$^{-1}$ together with *CI* and *VI* values before and after photodegradation for 12 h.

|  | $A_{1740}$ | $A_{1835}$ | *CI* | $A_{909}$ | $A_{2020}$ | *VI* |
| --- | --- | --- | --- | --- | --- | --- |
| **Initial** | -1.906 | -2.000 | 7.36 | -1.968 | -2.002 | 0.98 |
| **12 h photoreforming** | -1.947 | -1.997 | 3.95 | -1.971 | -1.997 | 0.99 |

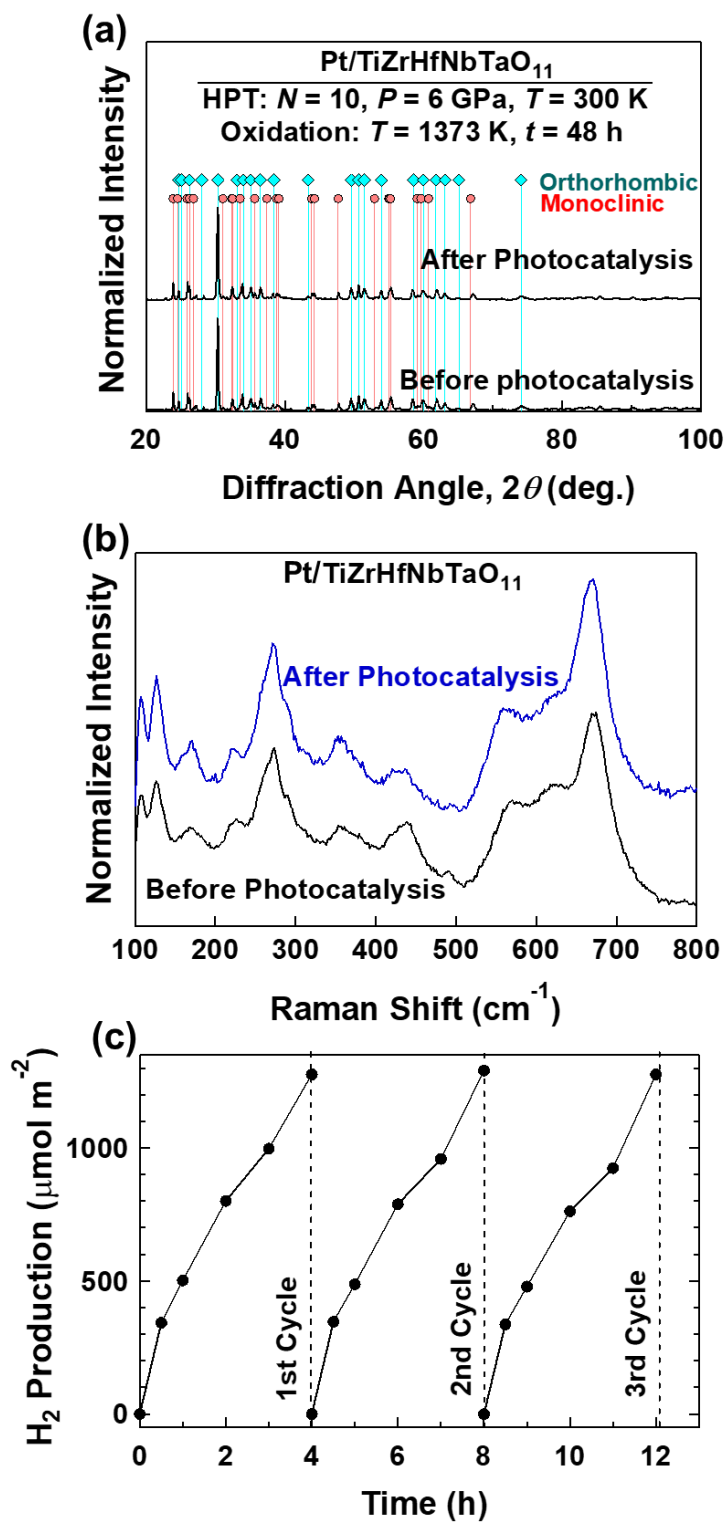

Figure 15. Stability and reusability of high-entropy oxide for photoreforming and simultaneous hydrogen production and PET plastic degradation. (a) XRD and (b) Raman spectra before and after catalytic test, and (c) $H_2$ production versus irradiation time for high-entropy oxide for three repeated tests.

## 4. Discussion

This study reports the first application of HEOs for simultaneous $H_2$ production and plastic degradation from the photoreforming process. $H_2$ production was observed from both PET microplastics and a real PET bottle using a HEO. $H_2$ production was confirmed by gas chromatography and plastic degradation is assessed during the photoreforming process by different methods. Firstly, degradation was confirmed by the decrease in mass and light transmittance, and the change in plastic surface morphology. Secondly, FTIR analysis indicated the deterioration of PET plastic and a decrease in the carbonyl index. Finally, the successful oxidation of PET plastic was verified by NMR measurement, indicating that PET is photodegraded to terephthalate, ethylene glycol, and formic acid. Three main questions need to be discussed here: (i) a comparison of the photoreforming performance of this HEO with the benchmark catalyst of P25 $TiO_2$ and other available catalysts from the literature, (ii) reasons for the superior catalytic activity of the HEO, and (iii) underlying mechanism for simultaneous $H_2$ production and PET degradation.

Regarding the first question, a comparison of the PET photoreforming performance of various catalysts is presented in Table 3. The catalytic activity of the HEO for $H_2$ production significantly surpasses that of the P25 $TiO_2$ benchmark catalyst (Fig. 8). The $H_2$ production yield after 4 h of light irradiation is 319 and 32 $\mu$mol m$^{-2}$ h$^{-1}$ for the HEO and P25 $TiO_2$, respectively. Although it is hard to compare the results reported in different publications due to the experimental differences in reactor, light source, catalyst concentration, etc., Table 3 suggests that the current HEO has one of the highest reported activities for photoreforming [3,14,17,18,67-70]. The $H_2$ production rate of the HEO is higher than many of the published results and comparable with highly active phosphides and sulfides. However, phosphides and sulfides are not as stable as oxides, particularly HEOs which are known for their high stability [28-30]. Besides the performance of the HEO for $H_2$ production, the second aspect that should be compared is plastic photo-oxidation. The degradation of different plastics using different catalysts is shown by the mass loss in Table 4. It can be concluded that this HEO is an active catalyst for plastic degradation compared to the data reported in the literature [71-79].

Regarding the second question, different aspects can be responsible for the high photocatalytic activity of the HEO: the presence of multiple cations in the catalyst leading to hybridized orbitals with higher activity for photocatalytic reactions [25,80], the coexistence of two phases preventing charge carrier recombination at interfaces (as heterojunctions) and boosting photocatalytic activity [40,48], the presence of lattice effects acting as activation sites of the catalyst [41,49], the high visible light absorbance (Fig. 6), the prolonged lifetime of charge carriers due to the defective nature of high-entropy materials (Fig. 7), and the lattice straining effect caused by the coexistence of five cations [28-30]. All these aspects contribute to the enhanced photocatalytic activity of HEOs as a new family of catalysts for photoreforming.

Regarding the third question, the mechanism of simultaneous catalytic $H_2$ production and plastic conversion is similar to the conventional photocatalytic $H_2$ evolution process with organic sacrificial agents such as methanol. In this process, PET plastic serves as a hole sacrificial agent, it is oxidized by the photogenerated holes, while the photoexcited electrons reduce water to produce $H_2$ [11-22]. When exposed to light irradiation, the high-entropy catalyst becomes photoexcited and its electrons migrate from the valence band to the conduction band. Under anaerobic conditions, these electrons actively participate in reducing $H^+$ of water to form $H_2$. Simultaneously, holes remain in the valence band of catalysts or generate reactive •OH radicals from $OH^-$ of water. These holes and radicals oxidize PET plastic into organic chemicals with smaller molecules. The reaction can be summarized as follows [11-22].

$$HEO + h\upsilon \rightarrow h^+_{(VB)} + e^-_{(CB)} \tag{7}$$

$$2H^+ + 2e^- \rightarrow H_2 \tag{8}$$

$$h^+_{(VB)} + OH^- \rightarrow \bullet OH \tag{9}$$

$$h^+_{(VB)} / \bullet OH + PET \rightarrow Organic\ Products \tag{10}$$

In summary, in photoreforming, not only plastic wastes are used as a sacrificial agent (instead of valuable chemicals like methanol), but also they are converted to value-added materials like hydrogen, terephthalate, ethylene glycol and formic acid. Therefore, the process has the potential to address two main issues of plastic waste management and clean $H_2$ fuel production [5,11-14]. While this study proves the high photoreforming performance of HEO micropowders synthesized by the assistance of severe plastic deformation [81], future experimental and theoretical studies and in situ observation are essential to synthesize nanopowders of these materials, to clarify their high catalytic activity and to understand the contribution of individual elements to their performance.

Table 3. Hydrogen production on high-entropy oxide from photoreforming of PET plastic, compared with P25 benchmark catalyst and reported data in the literature for various catalysts.

| Catalyst | Produced $H_2$ ($\mu mol\ m^{-2}\ h^{-1}$) | Ref. |
|---|---|---|
| High-entropy oxide | 319 | This study |
| P25 $TiO_2$ | 32 | This study |
| CN-CNTs-NiMo | 0.77 | [3] |
| CdS | 16.23 | [13] |
| Defect-rich $NiPS_3$ | 540.78 | |
| $CdS/d-NiPS_3$ | 926.48 | |
| CPDs-CN | 20.76 | [16] |
| $Pt/g-C_3N_4$ | 102 | [18] |
| $MXene/Zn_xC_{1-x}S$ | 123 | [67] |
| $O-CuIn_5S_8$ nanosheet | 27.39 | [68] |
| $CN_{0.14}$ porous microtube | 0.42 | [69] |
| $RP@Co_xP_y/Cd_{0.5}Zn_{0.5}S$ | 59.27 | [70] |

Table 4. Photoreforming efficiency of different catalyst for degradation of various plastics, including polyethylene terephthalate (PET), polyethylene (PE), high-density polyethylene (HDPE), low-density polyethylene (LDPE), polyethylene plastic-low density polyethylene (PP-LDPE), polyacrylonitrile (PAN), polyvinyl chloride (PVC), and polystyrene (PS).

| Type of plastic | Catalyst | Light source | Irradiation time (h) | Mass loss (%) | Ref. |
|---|---|---|---|---|---|
| PET | High-entropy oxide | UV | 16 | 100 | This study |
| PE film | $TiO_2$-MWCNTs | UV | 180 | 35 | [71] |
| HDPE microbeads | C, N- $TiO_2$ | Visible | 50 | 77.17 | [72] |
| LDPE | Fe-ZnO | Sunlight | 120 | 40 | [73] |
| PP-LDPE | $NiAl_2O_4$ | Visible | 5 | 10 | [74] |
| PE film | $TiO_2$ | Solar | 300 | 42 | [75] |
| PAN nanofibers | $TiO_2$/gCN | Simulated sunlight | 60 | 63.2 | [76] |
|  |  |  | 90 | 99 |  |
| PVC film | $TiO_2$/nanographite | UV | 30 | 17.24 | [77] |
| PS and PE | $TiO_2$ | UV | 24 | 99.99 | [78] |
| HDPE | Porous N-$TiO_2$ | Visible | 18 | 6.4 | [79] |

## 5. Conclusions

A high-entropy catalyst is synthesized by a three-step procedure of arc-melting, high-pressure torsion processing, and oxidation, and used for simultaneous $H_2$ production and PET plastic conversion through the photoreforming process. The synthesized high-entropy oxide showed better catalytic performance for $H_2$ generation and plastic degradation compared to the benchmark catalyst P25 $TiO_2$. In addition to green $H_2$, other valuable chemicals including terephthalate, ethylene glycol, and formic acid were also produced through the photoreforming using the high-entropy catalyst. This study introduces HEOs as a new family of catalysts for photoreforming which have a higher stability and comparable activity to highly active sulfides and phosphides.

**Acknowledgments**

This study is supported partly by Mitsui Chemicals, Inc., Japan, partly through Grants-in-Aid from the Japan Society for the Promotion of Science (JP22K18737), and partly by the ASPIRE project of the Japan Science and Technology Agency (JST) (JPMJAP2332).

**References**

[1] F. Zhang, M. Zeng, R.D. Yappert, J. Sun, Y.H. Lee, A.M. LaPointe, B. Peters, M.M. Abu-Omar, S.L. Scott, Polyethylene upcycling to long-chain alkylaromatics by tandem hydrogenolysis/aromatization, Science 370 (2020) 437-441.

[2] R. Geyer, J.R. Jambeck, K.L. Law, Production, use, and fate of all plastics ever made, Sci. Adv. 3 (2017) 1700782.

[3] X. Gong, F. Tong, F. Ma, Y. Zhang, P. Zhou, Z. Wang, Y. Liu, P. Wang, H. Cheng, Y. Dai, Z. Zheng, Photoreforming of plastic waste poly (ethylene terephthalate) via in-situ derived CN-CNTs-NiMo hybrids, Appl. Catal. B 307 (2022) 121143.

[4] T. Uekert, H. Kasap, F. Reisner, Photoreforming of nonrecyclable plastic waste over a carbon nitride/nickel phosphide catalyst, J. Am. Chem. Soc. 141(2019) 15201-15210.

[5] M. Du, Y. Zhang, S. Kang, X. Guo, Y. Ma, M. Xing, Y. Zhu, Y. Chai, B. Qiu, Trash to treasure: photoreforming of plastic waste into commodity chemicals and hydrogen over $MoS_2$-tipped CdS nanorods, ACS Catal. 12 (2022) 12823-12832.

[6] X. Jiang, H. Gong, Q. Liu, M. Song, C. Huang, In situ construction of NiSe/$Mn_{0.5}Cd_{0.5}S$ composites for enhanced photocatalytic hydrogen production under visible light, Appl. Catal. B 268 (2020) 118439.

[7] E.T. Edirisooriya, P.S. Senanayake, H.B. Wang, M.R. Talipov, P. Xu, H. Wang, Photo-reforming and degradation of waste plastics under UV and visible light for $H_2$ production using nanocomposite photocatalysts, J. Environ. Chem. Eng. 11 (2023) 109580.

[8] T. Kawai, T. Sakata, Photocatalytic hydrogen production from water by the decomposition of poly-vinylchloride, protein, algae, dead insects, and excrement, Chem. Lett. 10 (1981) 81-84.

[9] V. Preethi, S. Kanmani, Photocatalytic hydrogen production, Mater. Sci. Semicond. Process. 16 (2013) 561-575.

[10] B. Xia, Y. Zhang, B. Shi, J. Ran, K. Davey, S.Z. Qiao, Photocatalysts for hydrogen evolution coupled with production of value-added chemicals, Small Methods 4 (2020) 2000063.

[11] T. Uekert, M.F. Kuehnel, D.W. Wakerley, E. Reisner, Plastic waste as a feedstock for solar-driven $H_2$ generation, Energy Environ. Sci. 11 (2018) 2853-2857.

[12] H. Nagakawa, M. Nagata, Photoreforming of organic waste into hydrogen using a thermally radiative $CdO_x$/CdS/SiC photocatalyst, ACS Appl. Mater. Interfaces 13 (2021) 47511-47519.

[13] S. Zhang, H. Li, L. Wang, J. Liu, G. Liang, K. Davey, J. Ran, S.Z. Qiao, Boosted photoreforming of plastic waste via defect-rich $NiPS_3$ nanosheets, J. Am. Chem. Soc. 145 (2023), 6410-6419.

[14] C. Zhu, J. Wang, J. Lv, Y. Zhu, Q. Huang, C. Sun, Solar-driven reforming of waste polyester plastics into hydrogen over CdS/NiS catalyst, Int. J. Hydrogen Energy 51 (2024) 91-103.

[15] J.Q. Yan, D.W. Sun, J.H. Huang, Synergistic poly(lactic acid) photoreforming and $H_2$ generation over ternary $Ni_xCo_{1-x}$P/reduced graphene oxide/ g-$C_3N_4$ composite, Chemosphere 286 (2022) 131905.

[16] M. Han, S. Zhu, C. Xia, B. Yang, Photocatalytic upcycling of poly(ethylene terephthalate) plastic to high-value chemicals, Appl. Catal. B 316 (2022) 120764.

[17] Y.H. Lai, P.W. Yeh, M.J. Jhong, P.C. Chuang, Solar-driven hydrogen evolution in alkaline seawater over earth-abundant g-$C_3N_4$/$CuFeO_2$ heterojunction photocatalyst using microplastic as a feedstock, Chem. Eng. J. 475 (2023) 146413.

[18] T.K.A. Nguyen, T.P. Tran, X.M.C. Ta, T.N. Truong, J. Leverett, R. Daiyan, R. Amal, A. Tricoli, Understanding structure-activity relationship in Pt-loaded g-C$_3$N$_4$ for efficient solar-photoreforming of polyethylene terephthalate plastic and hydrogen production, Small Methods 8 (2023) 2300427.

[19] G. Peng, X. Qi, Q. Qu, X. Shao, L. Song, P. Du, J. Xiong, Photocatalytic degradation of PET microfibers and hydrogen evolution by Ni$_5$P$_4$/TiO$_2$/C NFs, Catal. Sci. Technol. 13 (2023) 5868-5879.

[20] Q. Wenbin, X. Qi, G. Peng, M. Wang, L. Song, P. Du, J. Xiong, An efficient and recyclable Ni$_2$P-Co$_2$P/ZrO$_2$/C nanofiber photocatalyst for the conversion of plastic waste into H$_2$ and valuable chemicals, J. Mater. Chem. C 11 (2023) 14359-14370.

[21] D. Jiang, H. Yuan, Z. Liu, Y. Chen, Y. Li, X. Zhang, G. Xue, H. Liu, X. Liu, L. Zhao, W. Zhou, Defect-anchored single-atom-layer Pt clusters on TiO$_{2-x}$/Ti for efficient hydrogen evolution via photothermal reforming plastics, Appl. Catal. B 339 (2023) 123081.

[22] J. Yuting, H. Zhang, L. Hong, J. Shao, B. Zhang, J. Yu, S. Chu, An integrated plasma-photocatalytic system for upcycling of polyolefin plastics, Chem. Sus. Chem. 16 (2023) e202300106.

[23] J. Qin, Y. Dou, F. Wu, Y. Yao, H.R. Andersen, C. Hélix-Nielsen, S.Y. Lim, W. Zhang, In-situ formation of Ag$_2$O in metal-organic framework for light-driven upcycling of microplastics coupled with hydrogen production, Appl. Catal. B 319 (2022) 121940.

[24] M.S. Lal, R. Sundara, High entropy oxides - a cost-effective catalyst for the growth of high yield carbon nanotubes and their energy applications, ACS Appl. Mater. Interfaces 11 (2019) 30846-30857.

[25] S.H. Albedwawi, A. AlJaberi, G.N. Haidemenopoulos, K. Polychronopoulou, High entropy oxides-exploring a paradigm of promising catalysts: a review, Mater. Des. 202 (2021) 109534.

[26] S. Akrami, P. Edalati, Y. Shundo, M. Watanabe, T. Ishihara, M. Fuji, K. Edalati, Significant CO$_2$ photoreduction on a high-entropy oxynitride, Chem. Eng. J. 449 (2022) 137800.

[27] P. Edalati, X.F. Shen, M. Watanabe, T. Ishihara, M. Arita, M. Fuji, K. Edalati, High-entropy oxynitride as low-bandgap and stable photocatalyst for hydrogen production, J. Mater. Chem. A 9 (2021) 15076-15086.

[28] A.J. Wright, Q. Wang, C Huang, A. Nieto, R. Chen, J. Luo, From high-entropy ceramics to compositionally-complex ceramics: a case study of fluorite oxides, J. Eur. Ceram. Soc. 40 (2020) 2120-2129.

[29] C. Oses, C. Toher, S. Curtarolo, High-entropy ceramics, Nat. Rev. Mater. 5 (2020) 295-309.

[30] S. Akrami, P. Edalai, M. Fuji, K. Edalati, High-entropy ceramics: review of principles, production and applications, Mater. Sci. Eng. R 146 (2021) 100644.

[31] A. Sarkar, L. Velasco, D.I. Wang, Q. Wang, G. Talasila, L. de Biasi, C. Kübel, T. Brezesinski, S.S. Bhattacharya, H. Hahn, B. Breitung, High entropy oxides for reversible energy storage, Nat. Commun. 9 (2018) 3400.

[32] T.X. Nguyen, J. Patra, J.K. Chang, J.M. Ting, High entropy spinel oxide nanoparticles for superior lithiation–delithiation performance, J. Mater. Chem. 8 (2020) 18963-18973.

[33] Y. Zheng, Y. Yi, M. Fan, H. Liu, X. Li, R. Zhang, M. Li, Z.A. Qiao, A high-entropy metal oxide as chemical anchor of polysulfide for lithium-sulfur batteries, Energy Storage Mater. 23 (2019) 678-683.

[34] G. Fang, J. Gao, J. Lv, H. Jia, H. Li, W. Liu, G. Xie, Z. Chen, Y. Huang, Q. Yuan, X. Liu, Multi-component nanoporous alloy/(oxy) hydroxide for bifunctional oxygen electrocatalysis and rechargeable Zn-air batteries, Appl. Catal. B 268 (2020) 118431.

[35] S. Zhou, Y. Pu, Q. Zhang, R. Shi, X. Guo, W. Wang, J. Ji, T. Wei, T. Ouyang, Microstructure and dielectric properties of high entropy Ba $(Zr_{0.2}Ti_{0.2}Sn_{0.2}Hf_{0.2}Me_{0.2})O_3$ perovskite oxides, Ceram. Int. 46 (2020) 7430-7437.

[36] R. Witte, A. Sarkar, R. Kruk, B. Eggert, R.A. Brand, H. Wende, H. Hahn, High-entropy oxides: An emerging prospect for magnetic rare-earth transition metal perovskites, Phys. Rev. Mater. 3 (2019) 034406.

[37] A.J. Wright, C. Huang, M.J. Walock, A. Ghoshal, M. Murugan, J. Luo, Sand corrosion, thermal expansion, and ablation of medium-and high-entropy compositionally complex fluorite oxides, J. Am. Ceram. Soc. 104 (2021) 448-462.

[38] H. Chen, W. Lin, Z. Zhang, K. Jie, D.R. Mullins, X. Sang, S.Z. Yang, C.J. Jafta, C.A. Bridges, X. Hu, R.R. Unocic, Mechanochemical synthesis of high entropy oxide materials under ambient conditions: dispersion of catalysts via entropy maximization, ACS Mater. Lett. 1 (2019) 83-88.

[39] P. Edalati, Q. Wang, H. Razavi-Khosroshahi, M. Fuji, T. Ishihara, K. Edalati, Photocatalytic hydrogen evolution on a high-entropy oxide, J. Mater. Chem. A 8 (2020) 3814-3821.

[40] P. Edalati, Y. Itagoe, H. Ishihara, T. Ishihara, H. Emami, M. Arita, M. Fuji, K. Edalati, Visible-light photocatalytic oxygen production on a high-entropy oxide with multiple-heterojunction introduction, J. Photochem. Photobio. A 433 (2022) 114167.

[41] S. Akrami, Y. Murakami, M. Watanabe, T. Ishihara, M. Arita, M. Fuji, K. Edalati, Defective high-entropy oxide photocatalyst with high activity for $CO_2$ conversion, Appl. Catal. B 303 (2022) 120896.

[42] K. Edalati, Z. Horita, A review on high-pressure torsion (HPT) from 1935 to 1988, J. Mater. Sci. Eng. A 652 (2016) 325-352.

[43] K. Edalati, A. Bachmaier, V.A. Beloshenko, Y. Beygelzimer, V.D. Blank, W.J. Botta, K. Bryła, J. Čížek, S. Divinski, N.A. Enikeev, Y. Estrin, G. Faraji, R.B. Figueiredo, M. Fuji, T. Furuta, T. Grosdidier, J. Gubicza, A. Hohenwarter, Z. Horita, J. Huot, Y. Ikoma, M. Janeček, M. Kawasaki, P. Krǎl, S. Kuramoto, T.G. Langdon, D.R. Leiva, V.I. Levitas, A. Mazilkin, M. Mito, H. Miyamoto, T. Nishizaki, R. Pippan, V.V. Popov, E.N. Popova, G. Purcek, O. Renk, Á. Révész, X. Sauvage, V. Sklenicka, W. Skrotzki, B.B. Straumal, S. Suwas, L.S. Toth, N. Tsuji, R.Z. Valiev, G. Wilde, M.J. Zehetbauer, X. Zhu, Nanomaterials by severe plastic deformation: review of historical developments and recent advances, Mater. Res. Lett. 10 (2022) 163-256.

[44] J.P. Couzinié, L. Lilensten, Y. Champion, G. Dirras, L. Perrière, I. Guillot, On the room temperature deformation mechanisms of a TiZrHfNbTa refractory high-entropy alloy, Mater. Sci. Eng. A 645 (2015) 255-263.

[45] O.N. Senkov, J.M. Scott, S.V. Senkova, D.B. Miracle, C.F. Woodward, Microstructure and room temperature properties of a high-entropy TaNbHfZrTi alloy, J. Alloys Compd. 509 (2011) 6043-6048.

[46] A.D. Wadsley, Mixed oxides of titanium and niobium. I., Acta Cryst. 14 (1961) 660-664.

[47] J. Galy, R.S. Roth, The crystal structure of $Nb_2Zr_6O_{17}$, J. Solid State Chem. 7 (1973) 277-285.

[48] K. Li, B. Peng, T. Peng, Recent advances in heterogeneous photocatalytic $CO_2$ conversion to solar fuels, ACS Catal. 6 (2016) 7485-7527.

[49] L. Ran, J. Hou, S. Cao, Z. Li, Y. Zhang, Y. Wu, B. Zhang, P. Zhai, L. Sun, Defect engineering of photocatalysts for solar energy conversion, Solar Rrl. 4 (2020) 1900487.

[50] M.R. Hoffmann, S.T. Martin, W. Choi, D.W. Bahnemann, Environmental applications of semiconductor photocatalysis, Chem. Rev. 95 (1995) 69-96.

[51] Z. Zhang, K. Liu, Z. Feng, Y. Bao, B. Dong, Hierarchical sheet-on-sheet $ZnIn_2S_4$/g-$C_3N_4$ heterostructure with highly efficient photocatalytic $H_2$ production based on photoinduced interfacial charge transfer, Sci. Rep. 6 (2016) 19221.

[52] M.C. Wu, C.H. Chen, W.K. Huang, K.C. Hsiao, T.H. Lin, S.H. Chan, P.Y. Wu, C.F. Lu, Y.H. Chang, T.F. Lin, K.H. Hsu, Improved solar-driven photocatalytic performance of highly crystalline hydrogenated $TiO_2$ nanofibers with core-shell structure, Sci. Rep. 7 (2017) 40896.

[53] K. Fujihara, S. Izumi, T. Ohno, M. Matsumura, Time-resolved photoluminescence of particulate $TiO_2$ photocatalysts suspended in aqueous solutions, J. Photochem. Photobiol. A 132 (2000) 99-104.

[54] W. Zhao, J. Liu, X. Wang, Y. Huang, H. Zhou, J. Yu, J. Liu, B. Hao, X. Wang, Y. Li, A novel high efficiency polar photocatalyst, Zn $(IO_3)_2$: synthesis, crystal structure and photocatalytic activity, New J. Chem. 45 (2021) 7844-7849.

[55] C.B. Hiragond, S. Biswas, N.S. Powar, J. Lee, E. Gong, H. Kim, H.S. Kim, J.W. Jung, C.H. Cho, B.M. Wong, S.I. In, Surface-modified Ag@ Ru-P25 for photocatalytic $CO_2$ conversion with high selectivity over $CH_4$ formation at the solid–gas interface, Carbon Energy 6 (2024) e386.

[56] K. Edalati, Q. Wang, H. Eguchi, H. Razavi-Khosroshahi, H. Emami, M. Yamauchi, M., Fuji, Z. Horita, Impact of $TiO_2$-II phase stabilized in anatase matrix by high-pressure torsion on electrocatalytic hydrogen production. Mat. Res. Lett. 2019 (7) 334-339

[57] T.T. Nguyen, K. Edalati, Impact of high-pressure columbite phase of titanium dioxide ($TiO_2$) on catalytic photoconversion of plastic waste and simultaneous hydrogen ($H_2$) production, J. Alloys Compd. 1008 (2024) 176722.

[58] M. Katai, P. Edalati, J. Hidalgo-Jimenez, Y. Shundo, T. Akbay, T. Ishihara, M. Arita, M. Fuji, K. Edalati, Black brookite rich in oxygen vacancies as an active photocatalyst for $CO_2$

conversion: experiments and first-principles calculations, J. Photochem. Photobio. A 449 (2024) 115409.

[59] Y. Shundo, T.T. Nguyen, S. Akrami, P. Edalati, Y. Itagoe, T. Ishihara, M. Arita, Q. Guo, M. Fuji, K. Edalati, Oxygen vacancy-rich high-pressure rocksalt phase of zinc oxide for enhanced photocatalytic hydrogen evolution, J. Colloid Interface Sci. 666 (2024) 22-34.

[60] T.T. Nguyen, K. Edalati, Brookite $TiO_2$ as an active photocatalyst for photoconversion of plastic wastes to acetic acid and simultaneous hydrogen production: comparison with anatase and rutile, Chemosphere 355 (2024) 141785.

[61] J. Hidalgo-Jiménez, T. Akbay, T. Ishihara, K. Edalati, Understanding high photocatalytic activity of the $TiO_2$ high-pressure columbite phase by experiments and first-principles calculations, J. Mater. Chem. A 11 (2023) 23523-23535.

[62] S. Chu, B. Zhang, X. Zhao, H.S. Soo, F. Wang, R. Xiao, H. Zhang, Photocatalytic conversion of plastic waste: from photodegradation to photosynthesis, Adv. Energy Mater. 12 (2022) 2200435.

[63] Y. Li, S. Wan, C. Lin, Y. Gao, Y. Lu, L. Wang, K. Zhang, Engineering of 2D/2D $MoS_2/Cd_xZn_{1-x}S$ photocatalyst for solar $H_2$ evolution coupled with degradation of plastic in alkaline solution, Solar Rrl. 5 (2021) 2000427.

[64] K.I. Ishibashi, A. Fujishima, T. Watanabe, K. Hashimoto, Detection of active oxidative species in $TiO_2$ photocatalysis using the fluorescence technique, Electrochem. Commun. 2 (2000) 207-210.

[65] T.T. Nguyen, K. Edalati, Impact of high-pressure torsion on hydrogen production from photodegradation of polypropylene plastic wastes, Int. J. Hydrog. Energy 81 (2024) 411-417.

[66] A.A. El-Saftawy, A. Elfalaky, M.S. Ragheb, S.G. Zakhary, Electron beam induced surface modifications of PET film, Radiat. Phys. Chem. 102 (2014) 96-102.

[67] B. Cao, S. Wan, Y. Wang, H. Guo, M. Ou, Q. Zhong, Highly-efficient visible-light-driven photocatalytic $H_2$ evolution integrated with microplastic degradation over $MXene/Zn_xCd_{1-x}S$ photocatalyst, J. Colloid Interface Sci. 605 (2022) 311-319.

[68] M. Du, M. Xing, W. Yuan, L. Zhang, T. Sun, T. Sheng, C. Zhou, Q. Qiu, Upgrading polyethylene terephthalate plastic into commodity chemicals paired with hydrogen evolution over a partially oxidized $CuIn_5S_8$ nanosheet photocatalyst, Green Chem. 25 (2023) 9818-9825.

[69] S. Guo, Y. Huang, D. Li, Z. Xie, Y. Jia, X. Wu, D. Xu, W. Shi, Visible-light-driven photoreforming of poly (ethylene terephthalate) plastics via carbon nitride porous microtubes, Chem. Commun. 59 (2023) 7791-7794.

[70] R. Li, F. Wang, F. Lv, P. Wang, X. Guo, J. Feng, D. Li, Y. Chen, Simultaneous hydrogen production and conversion of plastic wastes into valued chemicals over a Z-scheme photocatalyst, Int. J. Hydrogen Energy 51 (2024) 406-414.

[71] Y. An, J. Hou, Z. Liu, B. Peng, Enhanced solid-phase photocatalytic degradation of polyethylene by $TiO_2$–MWCNTs nanocomposites, Mater. Chem. Phys. 148 (2014) 387-394.

[72] M.C. Ariza-Tarazona, J.F. Villarreal-Chiu, J.M. Hernández-López, J.R. De la Rosa, V. Barbieri, C. Siligardi, E.I. Cedillo-González, Microplastic pollution reduction by a carbon and nitrogen-doped $TiO_2$: Effect of pH and temperature in the photocatalytic degradation process, J. Hazard. Mater. 395 (2020) 122632.

[73] S.M. Lam, J.C. Sin, H. Zeng, H. Lin, H. Li, Y.Y. Chai, M.K. Choong, A.R. Mohamed, Green synthesis of Fe-ZnO nanoparticles with improved sunlight photocatalytic performance for polyethylene film deterioration and bacterial inactivation, Mater. Sci. Semicond. Process. 123 (2021) 105574.

[74] C. Venkataramana, S.M. Botsa, P. Shyamala, R. Muralikrishna, R., Photocatalytic degradation of polyethylene plastics by $NiAl_2O_4$ spinels-synthesis and characterization, Chemosphere 265 (2021) 129021.

[75] X. u Zhao, Z. Li, Y. Chen, L. Shi, Y. Zhu, Solid-phase photocatalytic degradation of polyethylene plastic under UV and solar light irradiation, J. Mol. Catal. A. Chem. 268 (2007) 101-106.

[76] Y. Zhong, H. Chen, X. Chen, B. Zhang, W. Chen, W. Lu, Abiotic degradation behavior of polyacrylonitrile-based material filled with a composite of $TiO_2$ and $g-C_3N_4$ under solar illumination, Chemosphere 299 (2022) 134375.

[77] Y. Zhang, T. Sun, D. Zhang, Z. Shi, X. Zhang, C. Li, L. Wang, J. Song, Q. Lin, Enhanced photodegradability of PVC plastics film by codoping nano-graphite and $TiO_2$, Polym. Degrad. Stab. 181 (2020) 109332.

[78] I. Nabi, K. Li, H. Cheng, T. Wang, Y. Liu, S. Ajmal, Y. Yang, Y. Feng, L. Zhang, Complete photocatalytic mineralization of microplastic on $TiO_2$ nanoparticle film, Iscience 23 (2020) 101326.

[79] M.C. Ariza-Tarazona, J.F. Villarreal-Chiu, V. Barbieri, C. Siligardi, E.I. Cedillo-González, New strategy for microplastic degradation: Green photocatalysis using a protein-based porous N-$TiO_2$ semiconductor. Ceram. Int. 45 (2019) 9618-9624.

[80] J. Hidalgo-Jiménez, T. Akbay, T. Ishihara, K. Edalati, Investigation of a high-entropy oxide photocatalyst for hydrogen generation by first-principles calculations coupled with experiments: significance of electronegativity, Scr. Mater. 250 (2024) 116205.

[81] K. Edalati, A.Q. Ahmed, S. Akrami, K. Ameyama, V. Aptukov, R.N. Asfandiyarov, M. Ashida, V. Astanin, A. Bachmaier, V. Beloshenko, E.V. Bobruk, K. Bryła, J.M. Cabrera, A.P. Carvalho, N.Q. Chinh, I.C. Choi, R. Chulist, J.M. Cubero-Sesin, G. Davdian, M. Demirtas, S. Divinski, K. Durst, J. Dvorak, P. Edalati, S. Emura, N.A. Enikeev, G. Faraji, R.B. Figueiredo, R. Floriano, M. Fouladvind, D. Fruchart, M. Fuji, H. Fujiwara, M. Gajdics, D. Gheorghe, Ł. Gondek, J.E. González-Hernández, A. Gornakova, T. Grosdidier, J. Gubicza, D. Gunderov, L. He, O.F. Higuera, S. Hirosawa, A. Hohenwarter, Z. Horita, J. Horky, Y. Huang, J. Huot, Y. Ikoma, T. Ishihara, Y. Ivanisenko, J.I. Jang, A.M. Jorge Jr, M. Kawabata-Ota, M. Kawasaki, T. Khelfa, J. Kobayashi, L. Kommel, A. Korneva, P. Kral, N. Kudriashova, S. Kuramoto, T.G. Langdon, D.H. Lee, V.I. Levitas, C. Li, H.W. Li, Y. Li, Z. Li, H.J. Lin, K.D. Liss, Y. Liu, D.M. Marulanda Cardona, K. Matsuda, A. Mazilkin, Y. Mine, H. Miyamoto, S.C. Moon, T. Müller, J.A. Muñoz, M.Y. Murashkin, M. Naeem, M. Novelli, D. Olasz, R. Pippan, V.V. Popov, E.N. Popova, G. Purcek, P. de Rango, O. Renk, D.

Retraint, Á. Révész, V. Roche, P. Rodriguez-Calvillo, L. Romero-Resendiz, X. Sauvage, T. Sawaguchi, H. Sena, H. Shahmir, X. Shi, V. Sklenicka, W. Skrotzki, N. Skryabina, F. Staab, B. Straumal, Z. Sun, M. Szczerba, Y. Takizawa, Y. Tang, R.Z. Valiev, A. Vozniak, A. Voznyak, B. Wang, J.T. Wang, G. Wilde, F. Zhang, M. Zhang, P. Zhang, J. Zhou, X. Zhu, Y.T. Zhu, Severe plastic deformation for producing superfunctional ultrafine-grained and heterostructured materials: an interdisciplinary review, J. Alloys Compd. 1002 (2024) 174667.